\documentclass[sigplan,screen]{acmart}

\usepackage{cleveref}
\crefname{lstlisting}{Listing}{Listings}
\Crefname{lstlisting}{Listing}{Listings}

\usepackage{tikz}
\usetikzlibrary{positioning}
\usepackage{graphicx}

\usepackage[dvipsnames]{xcolor}
\definecolor{streamred}{HTML}{990000}
\definecolor{darkorange}{HTML}{E6550D}
\definecolor{tab0}{HTML}{3182BD}

\usepackage[table]{xcolor}
\definecolor{vlightgray}{gray}{0.87} 

\usepackage{pgfplots}
\usepackage{subcaption}
\usepackage{listings}

\usepackage{colortbl}
\usepackage{enumitem}
\usepackage{appendix}

\usepackage{pifont}
\newcommand{\yes}{\textcolor{green}{\ding{51}}} 
\newcommand{\no}{\textcolor{red}{\ding{55}}}
\newcommand{\stepops}[1]{#1}

\newcommand{\tileshape}[1]{\textcolor{gray}{\texttt{#1}}}

\newcommand{\streamshape}[1]{\textcolor{streamred}{\texttt{#1}}}

\newcommand{\minmem}{32}
\newcommand{\code}[1]{\texttt{\path{#1}}}

\newcommand{\tableop}[1]{\textbf{#1}}

\newcommand{\pluralstep}{Streaming Tensor Programs}

\AtBeginDocument{%
  }

\copyrightyear{2026}
\acmYear{2026}
\setcopyright{cc}
\setcctype{by}
\acmConference[ASPLOS '26]{Proceedings of the 31st ACM International Conference on Architectural Support for Programming Languages and Operating Systems, Volume 2}{March 22--26, 2026}{Pittsburgh, PA, USA}
\acmBooktitle{Proceedings of the 31st ACM International Conference on Architectural Support for Programming Languages and Operating Systems, Volume 2 (ASPLOS '26), March 22--26, 2026, Pittsburgh, PA, USA}
\acmDOI{10.1145/3779212.3790229}
\acmISBN{979-8-4007-2359-9/2026/03}

\pgfplotsset{compat=1.18} 
\begin{document}

\title{Streaming Tensor Programs: A Streaming Abstraction for Dynamic Parallelism}

\author{Gina Sohn}
\affiliation{
  \institution{Stanford University}
  \city{Stanford}
  \state{CA}
  \country{USA}
}
\email{ginasohn@stanford.edu}

\author{Genghan Zhang}
\affiliation{
  \institution{Stanford University}
  \city{Stanford}
  \state{CA}
  \country{USA}
}
\email{zgh23@stanford.edu}

\author{Konstantin Hossfeld}
\affiliation{
  \institution{Stanford University}
  \city{Stanford}
  \state{CA}
  \country{USA}
}
\email{hossfeld@stanford.edu}

\author{Jungwoo Kim}
\affiliation{
  \institution{Stanford University}
  \city{Stanford}
  \state{CA}
  \country{USA}
}
\email{jungwkim@stanford.edu}

\author{Nathan Sobotka}
\affiliation{
  \institution{Stanford University}
  \city{Stanford}
  \state{CA}
  \country{USA}
}
\email{nsobotka@stanford.edu}

\author{Nathan Zhang}
\affiliation{
  \institution{SambaNova Systems}
  \city{Palo Alto}
  \state{CA}
  \country{USA}
}
\email{stanfurd@stanford.edu}

\author{Olivia Hsu}
\affiliation{
  \institution{Stanford University}
  \city{Stanford}
  \state{CA}
  \country{USA}
}
\affiliation{
  \institution{Carnegie Mellon University}
  \city{Pittsburgh}
  \state{PA}
  \country{USA}
}
\email{owhsu@stanford.edu}

\author{Kunle Olukotun}
\affiliation{
  \institution{Stanford University}
  \city{Stanford}
  \state{CA}
  \country{USA}
}
\email{kunle@stanford.edu}

\renewcommand{\shortauthors}{Gina Sohn et al.}

\begin{abstract}

Dynamic behaviors are becoming prevalent in tensor applications, like machine learning, where many widely used models contain data-dependent tensor shapes and control flow. However, the limited expressiveness of prior programming abstractions for spatial dataflow accelerators (SDAs) forces these dynamic behaviors to be implemented statically and/or unoptimized.
To address these challenges, we present \pluralstep~(STeP), a streaming abstraction that enables dynamic tensor workloads to run efficiently on SDAs.
STeP introduces flexible routing operators, an explicit memory hierarchy, and symbolic-shape semantics that expose dynamic data rates and tensor dimensions. These capabilities unlock new optimizations, like dynamic tiling, dynamic parallelization, and configuration time-multiplexing, that adapt SDA execution to dynamic behaviors while preserving dataflow efficiency.
Using a cycle-approximate simulator on representative LLM layers and a full model with real-world traces, STeP enables: dynamic tiling that breaks the Pareto-optimal frontier from prior work, dynamic parallelization that improves latency by ${\sim}$2.72x, and configuration time-multiplexing that increases compute utilization by ${\sim}$2.64x over prior SDA abstractions and their implementations.

\end{abstract}


  
\maketitle 


\section{Introduction}\label{sec:intro}

The widespread use of compute- and memory-intensive tensor applications has increased the demand for performant hardware backends. This need for performance now drives the widespread adoption of high-throughput, highly parallel machines (like GPUs and dataflow architectures) for many tensor workloads, particularly large language models (LLMs)~\cite{vllm,sglang,deepspeed,megatronLM}. 
Under such circumstances, Spatial Dataflow Accelerators (SDAs)~\cite{revet,riptide,plasticine,sn40l,sn10,onyx} are emerging as a promising hardware architecture. SDAs are reconfigurable architectures composed of spatially distributed compute and memory units. By mapping computation onto a spatial fabric of pipelined compute and memory units, SDAs avoid several control overheads and enable aggressive operator fusion, pipelining, and fine-grained parallelism.
Prior work has empirically demonstrated that these architectural features enable SDAs to outperform state-of-the-art GPUs while delivering higher energy efficiency~\cite{kernellooping,chitty2024llm,intelligenceperwatt}.

\begin{table}[tbp]
\small
\setlength{\tabcolsep}{1.85pt}
\begin{tabular}{c!{\color{vlightgray}\vrule width 0.4pt}c!{\color{vlightgray}\vrule width 0.4pt}c!{\color{vlightgray}\vrule width 0.4pt}c!{\color{vlightgray}\vrule width 0.4pt}c!{\color{vlightgray}\vrule width 0.4pt}c}
\hline
Abstraction & 
\begin{tabular}[c]{@{}c@{}}Data\\Flow\end{tabular} & 
\begin{tabular}[c]{@{}c@{}c@{}}Explicit\\Data\\Rate\end{tabular} & 
\begin{tabular}[c]{@{}c@{}c@{}}Explicit\\Memory\\Hierarchy\end{tabular} &
\begin{tabular}[c]{@{}c@{}c@{}}Dynamic\\Routing\\\& Merging\end{tabular} & 
\begin{tabular}[c]{@{}c@{}}Dynamic\\On-chip\\ Tiling\end{tabular} \\
\hline
Spatial~\cite{spatial} & \no & \no & \yes  & \no & \no \\
Revet~\cite{revet}& \no & \no & \yes &  \yes\ (limited) & \no \\
StreamIt~\cite{streamit} & \yes & \yes & \no & \no & \no \\
SAM~\cite{sam} & \yes & \no & \no & \yes\ (limited) & \yes\ (limited) \\
Ripple~\cite{ripple} & \yes & \no & \no &  \yes & \no \\
STeP & \yes & \yes & \yes & \yes & \yes \\ \hline
\end{tabular}
\caption{Landscape of programming abstractions for SDAs}
\label{tab:prog-abs}
\end{table}

However, unlike the ample software support for static workloads on SDAs, existing SDA programming abstractions have limited support for accelerating dynamic workloads, as shown in \Cref{tab:prog-abs}.
The importance of supporting dynamic behaviors is increasing in many widely used tensor applications due to data-dependent tensor dimensions~\cite{splitwise,dynamollm,qwen2-vl,pang2025optimizingllminferencethroughput} and control flow~\cite{deepseekmoe,meta2025llama4,qwen3}.
Many such dynamic workloads can be characterized as asynchronously executing blocks with corresponding communication. This characterization aligns well with the execution model of SDAs, where compute and memory units run asynchronously and communicate via hardware FIFOs. As such, current SDA abstractions leave performance on the table even though the SDA hardware itself naturally maps well to these workloads. 

Most existing SDA programming abstractions fall into either imperative or dataflow-based designs. While imperative abstractions~\cite{spatial, sara, revet, riptide} offer high generality, 
dataflow designs, such as StreamIt~\cite{streamit}, SAM~\cite{sam}, and Ripple~\cite{ripple}, have emerged as they align better with the hardware's execution model.
However, they do not model an explicit memory hierarchy, and many were designed for a specific domain, limiting their ability to capture the broader range of dynamic tensor workloads.
SAM is limited to sparse tensor algebra kernels, and StreamIt adopts a synchronous dataflow model, making it challenging to express dynamic behaviors. Ripple adopts a design based on asynchronous blocks that can contain arbitrary imperative code. However, Ripple leaves the memory hierarchy implicit, making it difficult to express and discover efficient implementations of many important applications whose performance is dominated by data movement across the memory hierarchy~\cite{meta2025llama4,openai2025gptoss120bgptoss20bmodel,deepseekmoe,glm4.5,attention,qwen3}. Furthermore, opaque data rates at the abstraction level require lifting the imperative code within each asynchronous block to analyze the program in terms of data rates.

To address the limitations of prior SDA dataflow abstractions in expressing and optimizing dynamism, we propose \pluralstep~(STeP), a new streaming abstraction for accelerating dynamic tensor applications on SDAs.
STeP expresses data as streams, where tiles and buffers in the stream can have dynamic shapes. 
It consists of asynchronous dataflow blocks that provide three key properties: explicit memory hierarchy, symbolic data consumption and production rate, and data-dependent control flow operators.

These properties give STeP unique capabilities that are unavailable in prior abstractions for SDAs. First, STeP captures performance-critical metrics such as off-chip traffic, on-chip memory requirement, and operational intensity at the abstraction level. We show how STeP provides insight into memory-bound tensor applications and validate the captured metrics with a cycle-accurate simulator (\Cref{sec:implementation}).
STeP also enables expressing optimizations such as dynamic tiling, configuration time-multiplexing, and dynamic parallelization (\Cref{sec:evaluation}), which are not expressible in prior abstractions for SDAs. We evaluate each optimization on representative layers from open-source LLMs with real-world traces using a cycle-approximate simulator. 
Our evaluation shows that these optimizations break the Pareto-optimal frontier from prior work by delivering speedups and/or resource savings. Specifically, dynamic tiling delivers a Pareto Improvement Distance~\cite{ferretti2020,meng2016,fist2020}\footnote{The Pareto Improvement Distance measures the distance from a new design point $p$ to a reference Pareto frontier $P$. For more detail, see \Cref{subsec:dyn-tiling}.} of $1.33\times$${\sim}$$2.11\times$; Configuration time-multiplexing delivers $2.51\times$${\sim}$$2.64\times$ higher compute utilization; Dynamic parallelization achieves $1.14\times$${\sim}$ $2.72\times$ speedup.
We also evaluate the optimizations on end-to-end models, achieving upto $1.27\times$ speedup while using 69\% less on-chip memory and 54\% fewer compute resources on Qwen3-30B-A3B.
Lastly, we discuss future compilation to STeP and approaches for supporting the dynamic features of STeP in SDA hardware (\Cref{sec:discussion}).

Overall, our contributions are:
\begin{itemize}
    \item An asynchronous dataflow abstraction for SDAs (STeP) with first-class support for dynamism (\Cref{sec:step}).
    \item A symbolic system based on STeP's shape semantics to extract performance-critical metrics (\Cref{sec:implementation}).
    \item Optimizations that exploit the dynamic features and explicit memory hierarchy of STeP (\Cref{sec:evaluation}) and an outline of how those abstract dynamic features would be supported in SDA hardware (\Cref{sec:discussion}).
    \item A performance and resource utilization investigation on the impact of dynamic optimizations enabled by STeP on representative LLM applications (\Cref{sec:evaluation}). 
\end{itemize}

\section{Background} \label{sec:background}

This section provides background on the application, hardware, and programming abstractions discussed in this paper. 

\subsection{Dynamism in Machine Learning (ML)}\label{subsec:dynamism}

Although dynamism appears in many tensor applications, we will use ML workloads to illustrate real-world examples of dynamic behavior throughout this paper. Modern ML models exhibit diverse forms of dynamism and represent one of the most widely used tensor applications. ML workloads also demand high-throughput hardware backends, making them a primary driver for accelerators.

A prominent source of dynamism in recent ML workloads is the heavy use of data-dependent control flow. 
Mixture-of-Experts (MoE) is a model architecture where a subset of parameters, 
called experts, are activated for each input activation.
With every top-ranked open-source model now adopting the MoE architecture
~\cite{gemini2.5,qwen3,kimiK2,deepseekR1,deepseekV3,glm4.5,deepseekV3,hunyuanTurbos},\footnote{According to \url{https://lmarena.ai/leaderboard}, accessed on Nov. 12, 2025.} efficiently handling  
such control flow has become increasingly important.
ML workloads also frequently exhibit dynamic/ragged tensor shapes, driven by runtime parameters such as the number of requests, input resolution, and sequence length~\cite{dynamollm,qwen2-vl,splitwise,distserve}. Data-dependent control flow further amplifies this by making expert input shapes data-dependent.

\subsection{Spatial Dataflow Accelerators}\label{subsec:sda}
Spatial dataflow accelerators~\cite{plasticine,aurochs,revet,riptide,onyx,gorgon} are programmable architectures with spatially distributed hardware resources. A typical SDA consists of an array of reconfigurable compute units and memory units that communicate via hardware FIFOs and a network-on-chip. 
Instead of executing a sequential instruction stream as in the von Neumann model, SDAs represent programs as dataflow graphs, where nodes denote operations and edges represent explicit data dependencies. 
The nodes in a dataflow program graph are mapped to distributed compute and memory units, and the edges are mapped to hardware FIFOs and network-on-chip. 
The storage in SDAs is organized into multiple tiers, such as local PE storage, on-chip memory units, and off-chip memory. 
Most SDAs rely on the compiler or runtime to explicitly orchestrate and schedule the data movement from one storage tier to another~\cite{spatial,hsu2025stardust}.

\begin{figure}[t]
    \centering
    \includegraphics[width=\columnwidth]{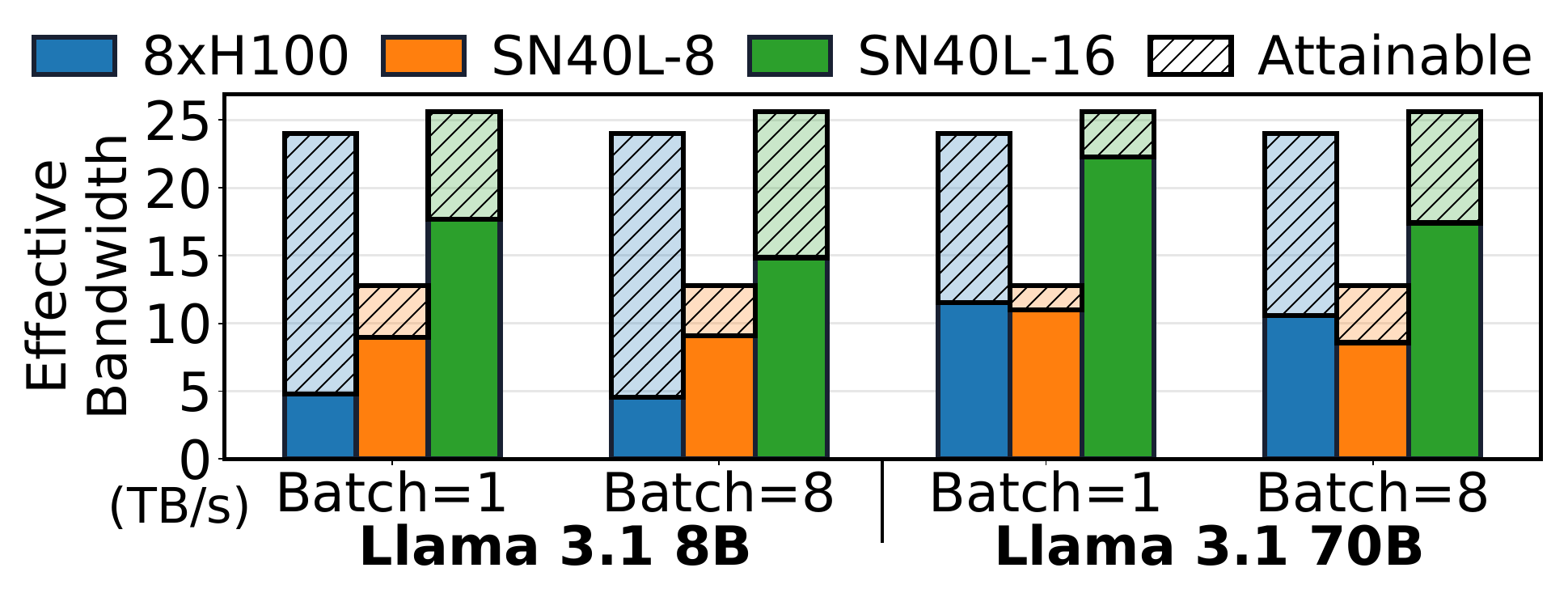}
    \caption{Comparison of SDAs versus GPUs. We express the effective bandwidth of each platform as solid bars, with slashed background bars indicating its peak HBM bandwidth. Effective bandwidth  is calculated using Roofline modeling and the percentage of peak throughput reported by~\cite{kernellooping}. \protect\footnotemark}
    \label{fig:gpu-sda-throughput}
    \Description{Four grouped bars to represent two models (Llama 3.1 8B, Llama 3.1 70B) and two batch sizes (1 and 8). Each grouped bar shows the peak HBM bandwidth and the effective HBM bandwidth for three platforms: eight H100 GPUs, eight SN40L, and 16 SN40L.}
\end{figure}

These architectural features allow SDAs to avoid the instruction decode, cache hierarchy, and control-flow divergence overheads of general-purpose processors and GPUs. Their dataflow execution model further enables aggressive operator fusion, pipelining, and fine-grained parallelism, which reduces off-chip memory traffic and synchronization overhead. Prior work shows that these advantages translate into both higher performance and energy efficiency compared to state-of-the-art GPUs~\cite{kernellooping,chitty2024llm,intelligenceperwatt}. For example, GPUs utilize less than half of their peak HBM bandwidth on Llama-3.1-8B and Llama-3.1-70B workloads (\Cref{fig:gpu-sda-throughput}), whereas the SN40L~\cite{sn40l}—a recent commodity SDA—achieves a higher fraction of peak HBM bandwidth during token generation. As this phase is heavily memory-bound, SN40L attains up to a $2\times$ speedup with half the peak HBM bandwidth (SN40L-8) and a $3.7\times$ speedup with comparable bandwidth (SN40L-16)~\cite{kernellooping,chitty2024llm}. Beyond throughput, SDAs also deliver higher task-level energy efficiency: the SN40L achieves $3.8\times$ and $4.6\times$ higher \emph{Intelligence per Joule} than NVIDIA B200 GPUs for Qwen3-32B and GPT-OSS-120B, respectively~\cite{intelligenceperwatt}.

\footnotetext{All models use a sequence length of 4K. 
GPU numbers were obtained by executing the models using TensorRT-LLM. The figure is reproduced with numbers from prior work with the original authors’ permission~\cite{kernellooping}.}

\subsection{Programming Abstractions for SDAs}\label{subsec:prog-abs}
SDAs can be programmed with either an imperative~\cite{spatial, revet, riptide} or a dataflow programming abstraction~\cite{sam,streamit, ripple} as listed in \Cref{tab:prog-abs}. While imperative abstractions offer high generality, 
they enforce a sequential instruction
order, which makes it challenging to exploit the inherent parallelism in the application~\cite{sara}. Furthermore, they lack explicit
primitives for asynchronous execution or queueing, which are crucial for optimizing dynamic workloads.

\begin{description}[leftmargin=8pt]
    \item[Spatial~\cite{spatial}] is an imperative programming abstraction for FPGAs and SDAs. It uses nested loops and provides explicit control over the memory hierarchy. However, control flow is only permitted in restricted regions of the program, and all memory constructs must be statically sized. Furthermore, transforming imperative loops into dataflow graphs that can be mapped to hardware introduces complexity in the compiler~\cite{sara}, and potentially results in suboptimal schedules.
    \item[Revet~\cite{revet}] is an imperative programming abstraction and compiler for expressing irregular applications on SDAs. It supports more flexible data-dependent control flow than Spatial via new dynamic primitives. However, its Dataflow Thread model restricts these primitives to scalars, which limits data reuse and prevents vectorized or tiled computation. 
    As a result, many large, memory-bound tensor applications in Revet are forced to use only static primitives to achieve high performance.
    Revet also cannot dynamically group scalar streams into dynamically sized tiles, making it unable to express optimizations that combine data-dependent control flow with dynamically-sized tiles.
\end{description}
    
Dataflow abstractions address these limitations with built-in support for dataflow and queueing. However, prior work either focuses only on a specific domain or lacks visibility and control over performance-critical decisions in many dynamic tensor applications. Throughout the paper, \emph{asynchronous dataflow} refers to an execution model in which dataflow blocks execute without global synchronization, and each block may exhibit dynamic data rates and latencies.

\begin{description}[leftmargin=8pt]
    \item[StreamIt~\cite{streamit}] is a synchronous dataflow abstraction used to map stream applications. It is not an abstraction dedicated to SDAs and can be used to target various streaming backends. Each node in the program graph has a fixed rate for consuming and producing data in the stream. While this design enables optimizations based on known data rates, this limits its ability to capture dynamic applications.
    
    \item[SAM~\cite{sam}] is the first asynchronous streaming tensor abstraction for SDAs. It introduces a clean dataflow model with primitives that can express the full space of sparse tensor algebra computations as streaming dataflow graphs. However, SAM is limited to sparse tensor operators, making it well-suited for exploring sparse workloads but not for dense dynamic tensor applications.

    \item[Ripple~\cite{ripple}] is an asynchronous dataflow abstraction and architecture that expresses the asynchronous pipeline parallelism enabled by SDAs. It has an implicit memory hierarchy and offers high generality by representing programs as asynchronous blocks that can contain any imperative code. While this model is sufficient for graph analytics and sparse workloads with inherently low reuse, dynamic tensor applications, such as dense ML, exhibit high temporal and spatial reuse. Therefore, visibility and explicit control over data movement across the memory hierarchy are crucial for performance; the lack of such control makes it challenging to analyze and express efficient schedules.
    Furthermore, its design makes the data rates of each asynchronous block opaque, requiring the compiler to infer them from the imperative code.

\end{description}

\section{\pluralstep}
\label{sec:step}
\pluralstep~(STeP) is a streaming abstraction for dynamic applications running on SDAs. In this section, we describe its stream representation and operators. 

\subsection{Stream-centric Design}\label{subsec:stream}
As an asynchronous dataflow model, STeP uses streams as the primary representation for data. Each stream has a compile-time determined rank and data type. 

\paragraph{Data Type} The data type of a stream can either be a tile, a selector, a reference to on-chip memory, or a tuple of these data types.
A tile is a two-dimensional regular matrix. STeP allows tiles to have dynamically defined shapes. Supporting dynamically-sized tiles is crucial for maximizing data reuse without excessive on-chip memory requirements when tiling tensors with runtime-determined shapes.
A selector is a multi-hot vector, which can express various routing and merging operators to support control flow (\Cref{subsubsec:route-merge}). STeP also enables read-only reference (i.e. addresses) to on-chip memory as a stream data type (\Cref{subsubsec:on-chip}).
The flexibility in data type enables lowering STeP to a broader range of SDAs more easily. For example, when the stream data type is restricted to only scalars, it cannot be directly mapped to SDAs with tiled processing units like systolic arrays without complicated lifting (e.g.  auto-vectorization).

\paragraph{Stop Tokens} STeP streams are logically equivalent to zero or more tensors. 
STeP streams embed the logical structure of the corresponding tensor into the data stream through \emph{stop tokens}.
STeP uses a similar stop token design to that of SAM~\cite{sam} as it was designed for asynchronous dataflow abstractions and allows for dimensions to be dynamic. The end of each dimension of the corresponding tensor is annotated with a stop token $S_N (N\geq1)$, where $N$ denotes the rank of that dimension (e.g. $N=1$ denotes the end of a vector) and at the end of multiple dimensions, STeP only emits the highest-level stop token. The \emph{Done} token ($D$) at the end indicates stream termination.

\paragraph{Stream Shape} 
The logical correspondence between a tensor and a STeP stream provides a foundation for defining shape semantics for streams. These semantics enable analyses and optimizations and improve debuggability by exposing dataflow block behaviors at the tensor level. 
Unlike the shape semantics of streams in synchronous dataflow~\cite{streamit}, which are straightforward due to fixed data rates, the shape semantics in asynchronous dataflow require careful design.

Each STeP stream has a \emph{rank} which is determined by the dimensionality of the corresponding tensor(s) in the stream. A rank-$N$ stream with a data type \texttt{T} is a stream of zero or more N-dimensional tensors of \texttt{T} and has a shape \textcolor{streamred}{$[D_N,\cdots,D_1,D_0]$}. Throughout the remainder of this section, we will use \textcolor{streamred}{red}, \textcolor{gray}{gray}, and black to denote the shape of the stream, the stream’s data type, and the tensor, respectively. 
STeP allows each \textcolor{streamred}{$D_i$} to be either a static-regular, a dynamic-regular, or a ragged dimension. 
A dimension is \emph{dynamic-regular} if its shape is a data-dependent constant. A dimension is \emph{ragged} if its shape can be various values (e.g. inner-most dimension in example \ref{eq:stream-shape-example}). A ragged dimension may be either dynamic or static, depending on whether its set of values is data-dependent. We will refer to either dynamic-regular or dynamic-ragged dimensions as \emph{dynamic dimensions}. The shape of dynamic-regular and ragged dimensions are expressed as equations and symbols (e.g. \textcolor{streamred}{$D_0$} in example \ref{eq:stream-shape-example}).
\begin{equation}
\label{eq:stream-shape-example}
\begin{tikzpicture}
\node (stream) {$1, 2, S_1,\ 3, S_2, \ \ \ 4, S_1,\ 5, 6, 7, S_2,\ D$};
\draw[<-] (stream.south west) -- (stream.south east);
\node[right=0.2cm of stream] (array) {\textcolor{streamred}{Shape: $[2, 2, D_0]$}};
\end{tikzpicture}
\end{equation}

Ragged dimensions have an absorbing property in the equations. If a dimension's shape equation contains a ragged dimension, that dimension will be treated as a new ragged dimension. For instance, when the inner-two dimensions are flattened in example (\ref{eq:stream-shape-example}), the resulting stream shape is \textcolor{streamred}{$[2,D_0']$} instead of \textcolor{streamred}{$[2,2\times D_0]$} where \textcolor{streamred}{$D_0'$} is a newly introduced symbol for the new ragged dimension. 

Certain STeP operators have restrictions on stream or data type shapes. Regular dimensions are more constrained than ragged dimensions, and static dimensions are more constrained than dynamic dimensions.\footnote{Regularity and data-dependence are orthogonal.}
Thus, if the operator accepts a dimension of a certain type, it also accepts more restrictive dimension types.

\subsection{STeP Operators}\label{subsec:step-ops}

STeP’s operators fall into five categories: (1) off-chip memory operators that stream tiled tensors between off-chip and on-chip memory; (2) on-chip memory operators that convert between streams and on-chip buffers; (3) dynamic routing and merging operators that implement data-dependent control flow; (4) higher-order operators that apply functions over stream elements; and (5) shape operators that modify stop tokens to change the stream’s logical tensor structure.\footnote{Supplementary syntax and shape semantics are provided in the Appendix.} 

\subsubsection{Off-chip Memory Operators}\label{subsubsec:off-chip}
Off-chip memory operators express the interface between on-chip and off-chip memory. Coupled with shape semantics, the off-chip memory operators can capture metrics such as off-chip memory traffic and operational intensity, exposing performance-critical metrics to the programmer or compiler.

\begin{description}[leftmargin=8pt]
    \item[LinearOffChipLoad] 
    As shown in \Cref{fig:offchip-example}, 
    this operator loads an input tensor with a specific shape (\emph{in\_mem\_shape}) from off-chip memory to on-chip memory in tiles. 
    It supports affine reads of the stored tensor using the stride and the shape arguments.
    The operator takes in a reference stream that controls how the stored tensor is repeatedly read. The reference stream's shape can contain any type of dimensions (static/dynamic-regular, and ragged). For each element in the reference stream, an affine read over the tensor is triggered (the correspondence is shown with black bold lines in \Cref{fig:offchip-example}). Because the reference stream serves as a trigger, its contents do not matter.
    \begin{figure}[h]
        \centering
        \includegraphics[width=\columnwidth]{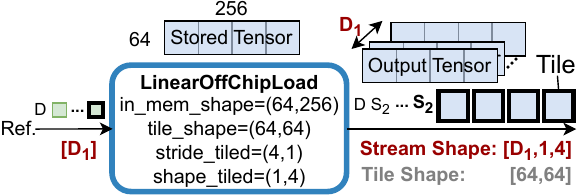}
        \caption{An example of a \stepops{LinearOffChipLoad} operator. The stored tensor is read in a row-major order $D_1$ times, where $D_1$ is the shape of a dynamic dimension. The stride and shape are expressed in terms of tiles. Therefore, the output stream shape is \textcolor{streamred}{$[D_1,64//64,256//64] = [D_1,1,4]$}. } 
        \Description{A figure of LinearOffchipLoad reading the stored tensor in row-major order $D_1$ times, where $D_1$ is the shape of a dynamic dimension.}
        \label{fig:offchip-example}
    \end{figure}

    \item[LinearOffChipStore] linearly stores the input stream's tiles to off-chip memory at the given address.
    
    \item[RandomOffChipLoad \& RandomOffChipStore] support random access to tensors stored in off-chip memory. Both take the base address, tile shape, and the in-memory tensor shape as arguments.
    As inputs, the \stepops{RandomOffChipLoad} operator has a read address stream, and the \stepops{RandomOffChipStore} has a write address stream and write data stream.
\end{description}

\subsubsection{On-chip Memory Operators} \label{subsubsec:on-chip}
These operators allow programs to leverage on-chip scratchpads and avoid off-chip memory accesses or recomputation. This expressiveness exposes a large design space of implementations that trade off on-chip memory usage against off-chip traffic. 

\begin{description}[leftmargin=8pt]
\item[Bufferize] stores portions of the stream to on-chip memory in linear order and outputs a stream of \emph{buffers} (a read-only reference to the allocated on-chip memory). The stream of buffers created by \stepops{Bufferize} becomes the input to \stepops{Streamify}, where any control-flow operator or shape operator (except Reshape) can be inserted between the two. The amount of data to be stored in on-chip memory is determined by the \emph{bufferize rank} argument. As shown in \Cref{fig:bufferize-example}, STeP allows the bufferized inner dimensions to be dynamic-regular dimensions and the outermost bufferize dimension can be a dynamic-ragged dimension.
    \begin{figure}[h]
        \centering
        \includegraphics[width=\columnwidth]{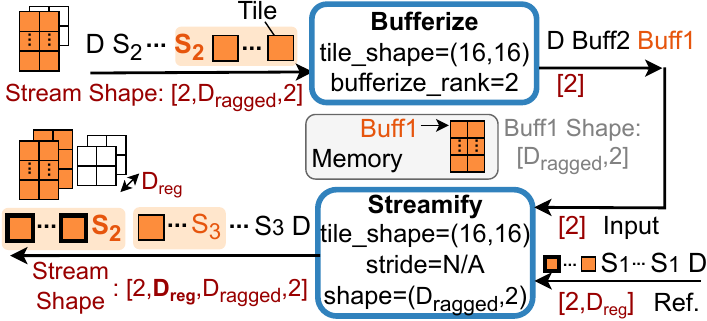}
        \caption{\stepops{Bufferize} stores the input stream tiles to on-chip memory until it sees a stop token larger than or equal to the bufferize rank. Then, a buffer is enqueued to the output stream, and the operator begins accumulating into a new on-chip memory (location).}
        \Description{A figure with a Bufferize and Streamify pair. The input stream has the shape [2, D_ragged, 2], which represents two buffers being streamed in. A buffer of shape [D_ragged,2] is created, and the content in each buffer is read using the Streamify operator. The Streamify takes a reference stream and reads each buffer a dynamic number of times.}
        \label{fig:bufferize-example}
    \end{figure}
    
\item[Streamify]
     supports reading data stored in on-chip memory by a dynamic number of times using a reference stream as shown in \Cref{fig:bufferize-example}. 
     When the buffer shape contains only static-regular dimensions, it supports affine reads over the buffer using its stride and shape arguments (similar to \stepops{LinearOffChipLoad}); otherwise, it linearly streams the tensor referenced by each buffer.
\end{description}

\subsubsection{Dynamic Routing and Merging Operators}
\label{subsubsec:route-merge}
These operators capture the essential routing and merging patterns to express data-dependent control flow and dynamic parallelism efficiently. STeP's dynamic routing and merging operators allow for flexible data types, while many prior SDA abstractions~\cite{spatial, revet, streamit} either lack support altogether or only support it under specific restrictions that significantly limit available parallelism.

\begin{figure}
    \centering
    \includegraphics[width=\columnwidth]{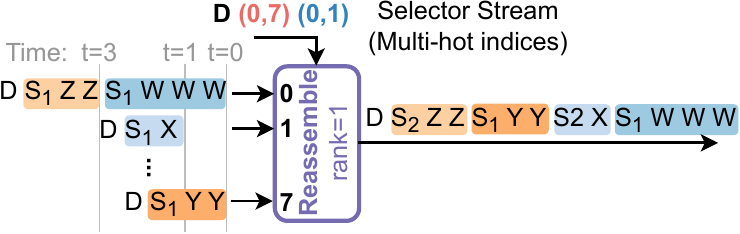}
    \caption{An example of \stepops{Reassemble}. The figure expresses multi-hot vectors in the selector stream as tuples of the indices of the nonzero elements. W-Z are data values.}
    \label{fig:reassemble}
    \Description{An reassemble (rank=1) takes in a selector stream of (0,1), (0,7) and eight input streams. The output collects the data from each input stream based on the selector's tuple.}
\end{figure}

\begin{description}[leftmargin=8pt]
    \item[Reassemble]
    merges data from many input streams based on the selector stream, as shown in~\Cref{fig:reassemble}. The input streams must all have the same rank, which is the \emph{reassemble rank}  \texttt{b} (In \Cref{fig:reassemble}, \texttt{b=1}). On every multi-hot vector in the selector stream, data up to the first $S_b$ from the selected input stream is merged to the output stream. When multiple input streams are selected by the multi-hot vector (selector), data is collected in the order the input is available. For the second multi-hot selector \textcolor{darkorange}{$(0,7)$} in \Cref{fig:reassemble}, data is collected from input stream 7 first. While routing inputs from one expert to the output stream, input streams don't get interleaved even though the other selected stream becomes available (e.g. in $t=1$). After collecting data from all selected input streams, the operator adds a new dimension by incrementing the stop token.
    \item[EagerMerge] is similar to \stepops{Reassemble} except that it collects data in the order they arrive. The operator has two output streams: the data stream and the selector stream, which denotes the index of the input stream from which each chunk of the stream was collected. For the input streams of \Cref{fig:reassemble}, EagerMerge will output the data stream in $W\rightarrow Y \rightarrow X \rightarrow Z$ or $Y\rightarrow W \rightarrow X \rightarrow Z$ order.
    \item[Partition] is the inverse of \stepops{Reassemble} and routes data up to the first $S_b$ from the input stream to the selected output streams ($b$ is the \emph{partition rank}).
\end{description}

\subsubsection{Higher-order Operators} \label{subsubsec:higher-order-ops}
These operators take a function supported by the hardware as an argument. 

 \begin{description}[leftmargin=8pt]

     \item[\stepops{Accum}] reduces over multiple inner dimensions of a stream. The operator takes the reduction rank, an initialization function, and an update function as arguments.
    \stepops{Accum} can express higher-order reductions by using an accumulator tile that is larger than the input tile. Similar to \Cref{fig:bufferize-example}, the accumulator for \stepops{Accum} can have a dynamic size.
    Together with \stepops{Bufferize}, this capability enables maximizing data reuse while using minimal on-chip memory when the application involves dynamically-sized tensors (we discuss related optimizations in \Cref{subsec:dyn-tiling}).
     
\item[Scan] is similar to \stepops{Accum} but emits the state of the accumulator on every input element. Therefore, the input and output streams have the same shape.

\item[Map] applies element-wise functions without changing the stream shape.

\item[FlatMap] expands each element in the stream to a stream of rank \texttt{b} by applying the supplied function. The resulting streams are concatenated into a single output stream. 
 \end{description}

\subsubsection{Shape Operators}\label{subsubsec:shape-ops}
These operators only modify stop tokens and do not alter the data contents of stream elements.
\begin{description}[leftmargin=8pt]
    \item[Flatten] takes the indices of two dimensions, which specify the range of dimensions that will be flattened. 
    \item[Reshape] splits a dimension into statically-sized chunks. When splitting the inner-most dimension, the operator takes in a padding value as an argument. The operator has two output streams: the data stream and the padding stream. The padding stream specifies whether each element in the output data stream was padded.
\item[Promote] adds a new outermost dimension to the input stream. 
Given an input stream's outermost dimension $D_a$, the new outermost dimension is $(1\ if\ (D_a > 0)\ else\ 0)$ to handle cases where the input stream is an empty stream.
\item[Expand] repeats each element in the input stream based on the reference stream as shown in \Cref{fig:expand-example}.
\begin{figure}[h]
        \centering
        \includegraphics[width=0.95\columnwidth]{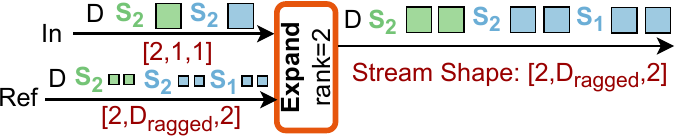}
        \caption{The \emph{expand rank} argument is set to the smallest stop token level of the input stream. The output stream has the same shape as the reference stream.}
        \label{fig:expand-example}
        \Description{A figure showing each element of the input stream being repeated based on the number of elements in the reference stream.}
    \end{figure}
\item[Zip] groups two streams with the same shape into a single stream with a tuple data type. 
\end{description}

\begin{figure*}[t]
\begin{center}
\centerline{\includegraphics[width=1.9\columnwidth]{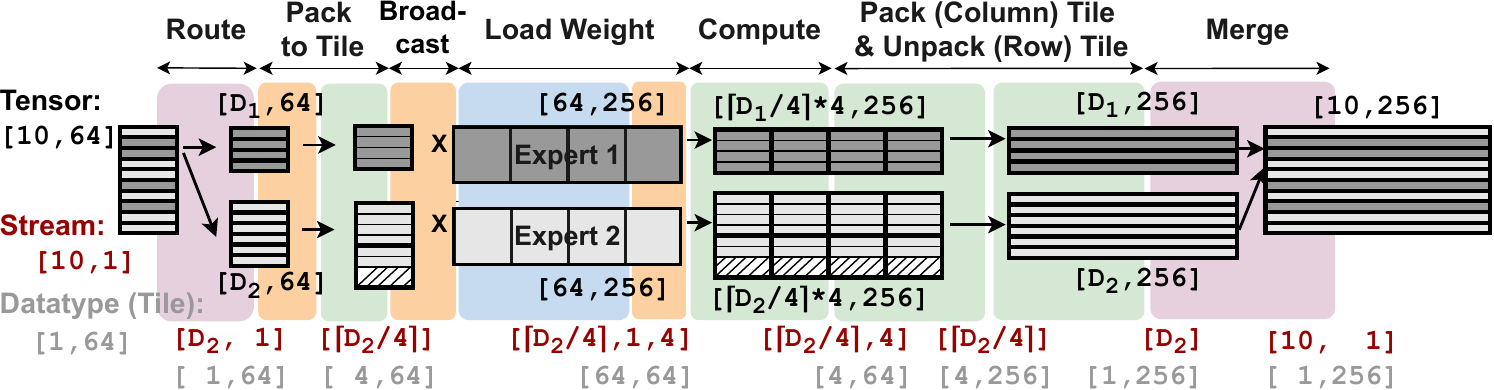}}
\caption{A simplified tensor-level MoE expressed example. Bold black lines mark tile boundaries that are streamed in row-major order. Black, dark red, and gray lists denote the shape of the tensor, stream, and the stream’s data type, respectively. 
}
\label{fig:big-example-tensor}
\Description{A figure showing how the tiled tensor is transformed throughout the simplified MoE program.}
\end{center}
\end{figure*}

\begin{figure*}[h]
    \centering
    \begin{tikzpicture}
        \node[inner sep=0pt] (img) at (0,0) {
            \begin{minipage}{0.45\textwidth}
                \centering
                \includegraphics[width=\textwidth]{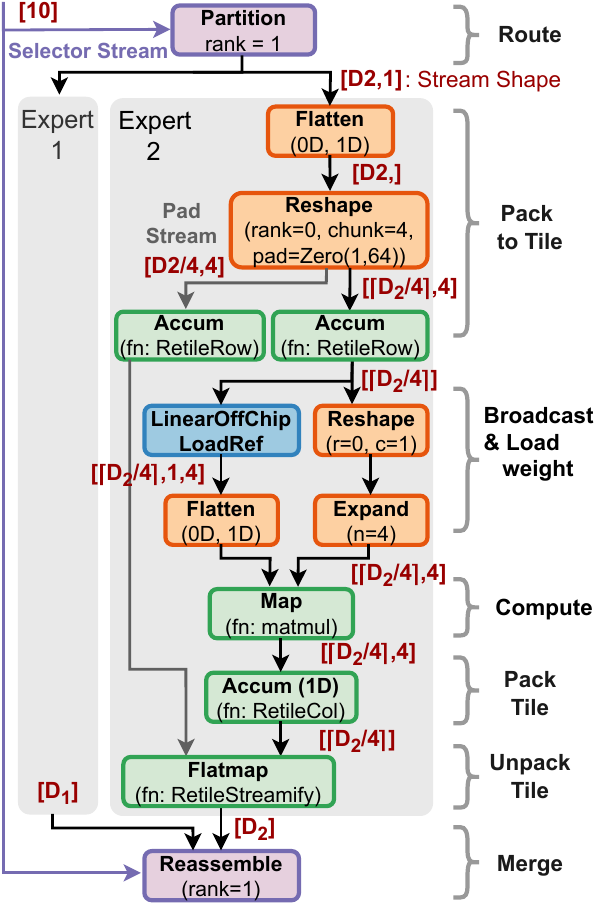}
                \captionof{figure}{The STeP graph for a simplified MoE. The background colors indicate corresponding regions in \Cref{fig:big-example-tensor}.}
                \label{fig:big-example-step}
                \Description{A STeP graph with each node colored to match the regions specified in \Cref{fig:big-example-tensor}.}
            \end{minipage}
        };
        
        \node[inner sep=6pt, anchor=west] (code) at ([xshift=1em]img.east) {
            \begin{minipage}{0.475\textwidth}
                \centering
                
                \tikz \node[draw, thick, rounded corners=2pt, inner sep=5pt] {
                    \begin{minipage}{\textwidth}
                        \begin{lstlisting}[language=Python, basicstyle=\small\ttfamily,
                        numbers=left,
    numberstyle=\tiny,
    stepnumber=1,
    numbersep=8pt,]
partition = Partition(in_stream, selector, 
    rank=1, num_consumers=N_ROUTED_EXPERTS)

expert_streams = []
for i in range(N_ROUTED_EXPERTS):
    flatten_in = Flatten((partition,i),0,1)
    reshape_to_tile = Reshape(...)
    collect_rows = Accum(...)
    collect_masks = Accum(...)
    weight_load = LinearOffChipLoadRef(
        ref=collect_rows,
        underlying=torch.randn(64,256), 
        stride=(4,1),
        shape=(1,4),
        tile_row=64, tile_col=256)
    flatten_w = Flatten(...)
    reshape_in = Reshape(...)
    expand_in = Expand(...)
    matmul_map = Map((expand_in, flatten_w),
        map_fn.Matmul(), compute_bw=1024)
    collect_cols = Accum(...)
    retile_streamify = Flatmap(...)
    expert_streams.append(retile_streamify)

output = Reassemble(expert_streams, selector, rank=1)
output.stream.shape = in_stream.stream.shape
print(output.stream.shape)
\end{lstlisting}
                    \end{minipage}
                };
                
                \captionof{lstlisting}{The code snippet for \Cref{fig:big-example-step} written in STeP's symbolic Python frontend.}
                \label{lst:code}
            \end{minipage}
        };
    \end{tikzpicture}
    \Description{A side-by-side figure in which the left shows \Cref{fig:big-example-step}, representing the STeP graph, and the right shows the corresponding code snippet written in the STeP symbolic Python frontend.}
\end{figure*}

\subsection{Putting it All Together: Simplified MoE}\label{subsec:moe-example}
To demonstrate how STeP combines to implement dynamic tensor applications, we will walk through a simplified MoE example. For this example, we use a simplified two-expert MoE layer, where each is a single matrix multiplication. Input rows are dynamically routed to one of the two branches, and their outputs are gathered back together after processing.
\Cref{fig:big-example-tensor} expresses the computation at the tensor level with tiling, and \Cref{fig:big-example-step} expresses the corresponding STeP graph. We will explain how the labelled operator regions in \Cref{fig:big-example-step} relate to the changes in the stream and the datatype shape. 
\begin{description}[leftmargin=8pt]
    \item[Route:] \stepops{Partition} takes in a \texttt{[10,64]} tensor, which is streamed as a \streamshape{[10,1]}-shaped stream of \tileshape{[1,64]} tiles. The output stream shapes are expressed symbolically as \streamshape{[$D_i$,1]} since each ($i$) branch receives a dynamic number of rows.
    \item[Pack to Tile:] With \tileshape{[1,64]} tile shapes, this region packs them into \tileshape{[4,64]} tiles to execute matrix-matrix multiplication rather than multiple matrix-vector operations. To do so, \stepops{Flatten} and \stepops{Reshape} first convert the stream shape from \streamshape{[$D_i$,1]} to \streamshape{[$\left\lceil D_i/4\right\rceil$,4]}. To pack a dynamic number of tiles in a stream into statically defined chunks, the \stepops{Reshape} operator pads the stream with the given pad value, which is \tileshape{[1,64]} zero-value tiles. The tiles in the stream are then packed into a larger tile using the \stepops{Accum} operator with the \texttt{RetileRow} function, which concatenates tiles row-wise.
    \item[Broadcast:] Since the matrix multiplication will be in inner-product dataflow order, each element in the input stream has to be broadcast by the number of tiles in the column dimension of the weight matrix. Therefore, we use \stepops{Reshape} and \stepops{Expand}\footnote{All STeP operators with an input reference stream have a static variant.} to do the stream shape conversion: \streamshape{[$\left\lceil D_i/4\right\rceil$]} $\rightarrow$ \streamshape{[$\left\lceil D_i/4\right\rceil$,1]} $\rightarrow$ \streamshape{[$\left\lceil D_i/4\right\rceil$,4]}.
    \item[Load weight:] The weight matrix is tiled along the column dimension and has to be loaded $\left\lceil D_i/4\right\rceil$ times. Since $D_i$ is dynamically determined, we must use a \stepops{LinearOffChipLoadRef} and feed the output stream of \stepops{Accum} (shape: \streamshape{[$\left\lceil D_i/4\right\rceil$]}) to its reference stream. This invokes reading the weight tensor $\left\lceil D_i/4 \right\rceil$ times as a \streamshape{[1,4]} stream of \tileshape{[64,64]} tiles.
    \item[Compute:] The matrix multiplication is done using a \stepops{Map} operator as we do not tile the reduction dimension.
    \item[Pack Tile, Unpack Tile, \& Merge:] To merge the streams in \tileshape{[1,256]} tile granularity, \stepops{Accum} first packs multiple tiles horizontally and then each tile is split row-wise into multiple smaller tiles using a \stepops{FlatMap} with the \stepops{RetileStreamify} function. Lastly, \stepops{Reassemble} gathers \tileshape{[1,256]} tiles based on a selector stream.
\end{description}

\section{Symbolic STeP Frontend and Performance Model} \label{sec:implementation}

In this section, we describe the programmability of STeP's symbolic frontend and how it combines with our simulator to capture performance-critical metrics.
We then present the performance model used in the STeP cycle-approximate simulator. 
We validate the metrics captured in the symbolic frontend and the simulator against a cycle-accurate hardware description language (HDL) simulation.
Lastly, this section contains a discussoin on how the symbolic frontend and simulator can be adapted to target different SDAs.

\subsection{Programmability of the Symbolic Frontend}
\label{subsec:programmability}
Continuing with the example in \Cref{subsec:moe-example}, \Cref{lst:code} is the equivalent code snippet for STeP's symbolic Python frontend.
Writing programs in STeP is similar to writing programs in PyTorch, but with schedules, such as parallelization, tiling, and memory placement. Instead of PyTorch operators, the programmer uses STeP operators, and the result of an operator is a stream instead of a tensor.
From the imperative coding perspective, writing programs in STeP means giving each instruction its own asynchronously executing loops connected via streams.

STeP's stream-centric design enables operator fusion by construction and eliminates complicated compiler passes for extracting parallelism from imperative code. However, as tensors between operators are expressed as streams, both the corresponding tensor shape and the stream shape (i.e., how the elements in the tensor are being streamed) have to align between operators. STeP's shape semantics enable the symbolic frontend to internally verify that stream shapes align between the producer and consumer, and allow programmers to inspect the stream shape (see line 27 of \Cref{lst:code}).
Stream shapes can also be used to exploit known program properties. For example, the new dynamic dimension introduced by \stepops{Reassemble} can be substituted with the input stream’s shape as shown in line 26 of \Cref{lst:code}.

STeP provides two sets of memory operators (off-chip and on-chip) with similar interfaces, allowing programmers to select operators based on the desired memory placement. For instance, line 10 of \Cref{lst:code} uses off-chip memory to load weights, but if those weights are already resident in on-chip memory, the operator can be replaced with \stepops{Streamify}. 
Such visibility and control over the memory hierarchy enable analyzing and exploring different tiling schemes, which are a crucial scheduling knob for many tensor applications.

\subsection{Analysis with Symbolic Shape Semantics}\label{subsec:python-frontend}
The symbolic frontend implements symbolic expressions for off-chip memory traffic and on-chip memory requirements for each operator using SymPy~\cite{sympy}.
The total off-chip memory traffic and on-chip memory requirement of a program is obtained by summing the expressions for every operator in the program graph.
Throughout this subsection, we use $||X||$ to denote the cardinality of a buffer or stream $X$, defined as the product of its dimension sizes. $dtype$ denotes data type, and $|x|$ to denote the size of a data type $x$. 

\paragraph{Off-chip Traffic} As the off-chip memory traffic only occurs in off-chip memory operators (\Cref{subsubsec:off-chip}) the equation for other operators is zero, and the equation for off-chip memory operators is:
\begin{equation*}
    ||output\ stream||\times|ouput\ stream\ dtype|
\end{equation*}

If the target SDA assumes that no other STeP operators spill to off-chip memory, the summed off-chip traffic equation represents the program's total off-chip memory traffic and can be used to compute operational intensity. Otherwise, it provides a lower bound on off-chip traffic and thus an upper bound on achievable operational intensity.

\paragraph{On-chip Memory Requirement}
In our simulator, we use the following equations for each operator. Other operators return zero because they can be fully streamed without materialization.
\begin{description}[leftmargin=10pt]
    \item[Off-chip memory operators:] $|output\ dtype|\times 2$
    \item[Bufferize:] $|input\ dtype|+||buffer||\times|input\ dtype|\times2$ \\
    We multiply by two, assuming double buffering. 
    \item[Accum, Scan, Expand:] $|output\ dtype|$
    \item[Map, Accum with matrix multiplication:]\ \\
    $(16\times in\_tile\_col + |weight\ tile| + |output\ tile|)$\\
    The $in\_tile\_col$ denotes the input tile’s innermost dimension size. The output tile size is included only for \stepops{Accum}. We account for the storage of partial-input tiles and the full weight tile because we use inner-product matrix multiplication. We multiply by 16 to mirror the decomposition of STeP-level tiles into $16\times16$ tiles that align with the hardware’s compute-unit tile size.\footnote{Refer to \Cref{subsec:hierarchical-tiling} for more detail on hierarchical tiling.}
\end{description}

\paragraph{Handling data dependencies}

When dynamic-regular or ragged dimensions are present, the metrics produced by the symbolic frontend include symbolic variables.\footnote{The symbolic frontend also introduces symbols for static ragged dimensions, as explicitly tracking all ragged values would significantly increase the complexity of metric computation.} By substituting these symbols with different input shapes or control-flow decisions, programmers can quickly analyze off-chip traffic and on-chip memory requirements.

The exact values for each metric are obtained by invoking the simulator explained in the following subsection. The symbolic frontend tracks operators whose metrics depend on runtime data and enables off-chip traffic or on-chip memory measurements for those operators during simulation. 
The values from the symbolic frontend and the simulator are then aggregated to produce the final concrete metrics.

\subsection{Performance Model for the Simulator}\label{subsec:simulator}

Since the symbolic STeP frontend has no timing information, we implement a simulator backend for STeP in Rust using the Dataflow Abstract Machine~\cite{dam} simulation framework.

To model the data transfer between off-chip memory and on-chip memory, the simulator implements an HBM node that emulates the timing behavior of Ramulator 2.0~\cite{ramulator2}, a cycle-accurate DRAM simulator.
The latency of accessing on-chip memory is factored into the higher-order operators that execute arithmetic functions, using Roofline modeling. Each higher-order operator is allocated a compute bandwidth (FLOPs/cycle). On each input element in the stream, the operators increment cycles based on the following equation:
\begin{equation*}
    \text{max}(\frac{size\ of\ inputs}{on\text{-}chip\ mem\ BW},\frac{total\ FLOPs}{compute\ BW},\frac{size\ of\ outputs}{on\text{-}chip\ mem\ BW})
\end{equation*}

As shown by the \texttt{matmul\_map} in \Cref{lst:code}, the $compute\ BW$ is provided by the programmer, and $total\ FLOPs$ is computed within the function supplied to the higher-order operators, as this value depends on the specific computation the function performs. The first and last entries in the equation are zero when the input and output are streamed directly between compute units via FIFOs.

\subsection{Portability}\label{subsec:portability}
As an abstraction, STeP is not tied to a specific hardware implementation~\cite{sam} and is portable across diverse SDA implementations with software-managed scratchpads~\cite{onyx,plasticine,aurochs,sn10,sn40l,revet}, which we discuss in \Cref{subsec:dynamic-hardware}. The equations in the symbolic frontend
can be customized to capture hardware-specific operator details, such as hardware tile sizes and matrix-multiplication implementation. 
 If performance bottlenecks shift, additional cost functions can be added to STeP operators to obtain performance-correlated metrics (e.g. on-chip traffic, compute).
In the simulator, operator initiation intervals and latencies can be adjusted to match hardware characteristics~\cite{dam}. For example, when multiple \stepops{Bufferize} operators share an on-chip memory unit, a scratchpad simulator can replace the Roofline model to capture on-chip memory contention. Also, different memory technologies can be modeled by reconfiguring or replacing the simulator’s HBM node~\cite{ramulator2}.\footnote{Detailed instructions for customizing the frontend and simulator can be found in the artifact repository described in \Cref{subsec:artifact-info}.}

\subsection{Validation}\label{subsec:validation}
We validate the simulator by comparing the performance to a cycle-accurate HDL simulation. We also compare the off-chip traffic captured in the symbolic STeP frontend with the performance to validate the usefulness of the metrics captured in the symbolic frontend.

\noindent\emph{\textbf{Workload and Hardware Model.}} We use a SwiGLU~\cite{swiglu} layer as the workload since it contains representative computations in ML models such as matrix multiplication, activation function, and row-wise reduction. We choose a spatial architecture of compute units that operate on $16\times16$ BFloat16 tiles, each having an initiation interval of one. We pair compute tiles with distributed on-chip memory units, each capable of reading and writing one tile per cycle. 

The HDL model is implemented in Bluespec SystemVerilog and executed in a cycle-accurate BlueSim simulator~\cite{nikhil2004bluespec, bourgeat2020bluespec}. Off-chip access delays are integrated using the Ramulator2 library calls using a configuration of an HBM2 subsystem with 8 stacks. The on-chip memory bandwidth is configured as 256 (bytes/cycle), and the cycle-approximate STeP simulator uses the same memory configurations. We measure the total execution time from the first off-chip read to the last off-chip write. 

\noindent\emph{\textbf{Mapping Methodology.}} 
We apply a graph transformation to partition tiles into smaller physical tiles that match the fabric's compute tile size.\footnote{An example transformation is shown in \Cref{fig:downtiling} in the Appendix.} After the transformation, every node in the graph maps to a dedicated unit in the HDL design, which we attach to a congestion-free interconnect. The programmer-specified compute bandwidth determines how many compute units are mapped to each STeP node.

\begin{figure}
    \centering
    \begin{tikzpicture}
\footnotesize
\definecolor{myblue}{rgb}{0.1215686,0.4666667,0.7058824}
\definecolor{myorange}{rgb}{1.0,0.4980392,0.05490196}

\begin{axis}[
    name=plot1,
    width=7.5cm,
    height=3cm,
    ymin=0, ymax=40000,
    ylabel={Clock Cycles},
    ylabel style={yshift=-5pt},
    xtick=data,
    yticklabel={\(\pgfmathprintnumber{\tick}\)},
    symbolic x coords={
        a, b, c, d, e, f, g, h, i, j, k, l, m, n, o
    },
    xticklabels={}, 
    xticklabel=\empty,
    ymajorgrids=true,
    xmajorgrids=false,
    legend style={at={(3.2cm,1.8cm)}, anchor=center, legend columns=-1},
    enlarge x limits=0.02,
]
\addplot+[only marks, mark=x, color=myblue, thick, mark options={scale=2}] coordinates {
    (a, 36513) (b, 30726) (c, 28805) (d, 27356) (e, 26496)
    (f, 22577) (g, 17923) (h, 16250) (i, 15015) (j, 13882)
    (k, 16524) (l, 11617) (m, 9700) (n, 9396) (o, 8805)
};
\addplot+[only marks, mark=+, color=myorange, thick, mark options={scale=2}] coordinates {
    (a, 33289) (b, 29401) (c, 27502) (d, 26991) (e, 27503)
    (f, 21224) (g, 17098) (h, 15288) (i, 14767) (j, 15292)
    (k, 15218) (l, 11186) (m, 9538) (n, 9275) (o, 11061)
};
\legend{HDL Implementation, STeP Simulator}
\end{axis}

\begin{axis}[
    at=(plot1.below south west), anchor=north west,
     ylabel near ticks, yticklabel pos=right,
    yshift=.1cm,
    width=7.5cm,
    height=3cm,
    ymin=0, ymax=10,
    ylabel={\shortstack{Off-chip Memory \\ Traffic (MB)}},
    ylabel style={yshift=3pt},
    xlabel={\shortstack{Tile Size \\ (Batch Dimension, Hidden Dimension, MoE Intermediate Dimension)}},
    xtick=data,
    xticklabel style={rotate=45, anchor=north east},
    symbolic x coords={
        a, b, c, d, e, f, g, h, i, j, k, l, m, n, o
    },
    yticklabel={\(\pgfmathprintnumber{\tick}\)},
    xticklabels={
        {${(16,256,16)}$}, {${(16,256,32)}$}, {${(16,256,64)}$},
        {${(16,256,128)}$}, {${(16,256,256)}$}, {${(32,256,16)}$},
        {${(32,256,32)}$}, {${(32,256,64)}$}, {${(32,256,128)}$},
        {${(32,256,256)}$}, {${(64,256,16)}$}, {${(64,256,32)}$},
        {${(64,256,64)}$}, {${(64,256,128)}$}, {${(64,256,256)}$}
    },
    ymajorgrids=true,
    xmajorgrids=false,
    enlarge x limits=0.02,
]
\addplot+[only marks, mark=x, color=myblue, thick, mark options={scale=2}] coordinates {
    (a, 8.481792) (b, 7.429120) (c, 6.902784) (d, 6.639616) (e, 6.508032)
    (f, 5.323776) (g, 4.271104) (h, 3.744768) (i, 3.481600) (j, 3.350016)
    (k, 3.744768) (l, 2.692096) (m, 2.165760) (n, 1.902592) (o, 1.771008)
};
\addplot+[only marks, mark=+, color=myorange, thick, mark options={scale=2}] coordinates {
    (a, 8.454144) (b, 7.405568) (c, 6.881280) (d, 6.619136) (e, 6.488064)
    (f, 5.308416) (g, 4.259840) (h, 3.735552) (i, 3.473408) (j, 3.342336)
    (k, 3.735552) (l, 2.686976) (m, 2.162688) (n, 1.900544) (o, 1.769472)
};
\end{axis}
\end{tikzpicture}
    \caption{Cycle-count and memory traffic comparison of a SwiGLU Layer with different tile sizes. The full sizes of the batch dimension, hidden dimension, and MoE intermediate dimension are 64, 256, and 512, respectively. 
    }
    \label{fig:step-vs-bluespec}
\Description{Two stacked plots comparing HDL Implementation (blue x) and STeP Simulator (orange plus) across 15 tile size configurations varying batch dimension (16, 32, 64), hidden dimension (256), and MoE intermediate dimension (16-256). Top plot shows clock cycles (0-40,000), bottom shows off-chip memory traffic (0-10 MB). Both metrics decrease with larger tile sizes, and the two implementations track closely with HDL showing slightly higher values.}
\end{figure}
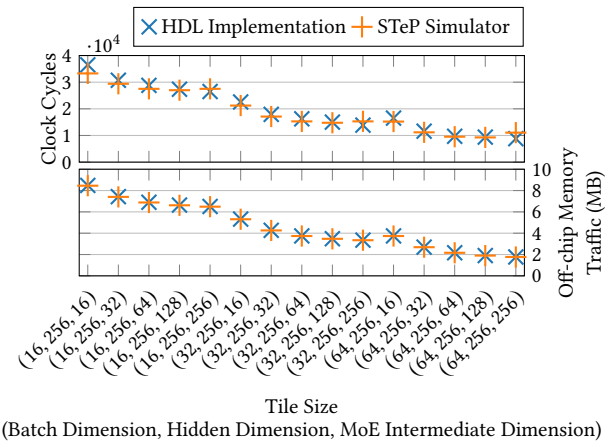

\noindent\emph{\textbf{Results.}} As shown in~\Cref{fig:step-vs-bluespec}, the STeP simulator's cycle-count closely matches that of the HDL simulator, with a Pearson correlation of $0.99$, when sweeping different tile sizes.
As the application is memory-bound in the given hardware configuration, decisions on data transfer across the memory hierarchy significantly impact performance, highlighting the importance of having visibility and control over these decisions in the abstraction. The high correspondence between the off-chip traffic captured in the symbolic STeP frontend and the HDL simulator's performance and incurred off-chip traffic suggests that the metrics captured in the symbolic frontend and the STeP simulator can provide valuable insights into the performance of a given STeP graph.

\section{Evaluation}\label{sec:evaluation}

In this section, we evaluate STeP’s ability to explore efficient schedules for dynamic ML models by implementing optimizations that were not expressible in prior abstractions for SDAs.
Key STeP features that enable each optimization are listed in \Cref{tab:optimizations}. 
Our evaluation shows that these optimizations enable Pareto-optimal design points over prior work and deliver speedups or resource savings.
To show how the optimizations integrate with full LLM inference, we also evaluate them on end-to-end models.

\subsection{Methodology}\label{subsec:eval-methodology}

\emph{\textbf{Workload.}} 
Our workloads consist of two layers: Mixture-of-Experts (MoE) with SwiGLU~\cite{swiglu} experts and attention~\cite{attention}.
We focus on these layers because 
 they dominate end-to-end inference latency. For example, when running DeepSeek-R1 on $64\times$ B200 GPUs, attention and MoE layers together account for approximately 80\% of the total latency~\cite{yun2025new}.
We use configurations from Qwen3-30B-A3B and Mixtral-8x7B. We choose Qwen3-30B-A3B because it shares a common architecture with many of the top-20 open-source models~\cite{qwen3,deepseekmoe,gemini2.5,kimiK2,deepseekR1,deepseekV3,glm4.5,hunyuanTurbos} on the LM Arena leaderboard~\cite{lmarena_leaderboard}. Although Mixtral is relatively older, we include it to demonstrate the impact of our optimizations across varied expert-activation patterns in MoE models.

\noindent\emph{\textbf{Dataset}}
For the attention layer, the KV cache length for each batch is sampled from the AzureLLMInference dataset \cite{splitwise}.
For the MoE layer, we use expert-routing data collected by running the models on the HH-RLHF~\cite{hh-rlhf} request traces.\footnote{More details on the datasets can be found in \Cref{subsec:dataset-appendix} of the Appendix.}

\noindent\emph{\textbf{Simulator Setup.}} We use the STeP simulator descibed in \Cref{subsec:simulator}. The bandwidth of each on-chip memory unit is set to 64 bytes/cycle, and the off-chip memory bandwidth is set to 1024 bytes/cycle, matching the configurations of recent reconfigurable dataflow accelerators~\cite{plasticine, sn40l}. 

\noindent\emph{\textbf{Baseline Design.}} 
We chose Revet~\cite{revet} as our baseline since it has the most extensive support for dynamic behaviors among SDA programming abstractions with explicit memory hierarchy.
However, as discussed in \Cref{subsec:prog-abs}, restrictions in Revet’s dataflow thread model and its lack of support for dynamically-sized tiles make it impossible to express the proposed optimizations.
Therefore, we use STeP to implement schedules that are expressible in Revet and treat these implementations as the baseline. 
All other scheduling and mapping decisions are identical between the baseline and optimized implementations.
We do not compare against Ripple as it targets SDAs without on-chip scratchpads (see \Cref{subsec:prog-abs}). 

\begin{table}[tbp]
\small 
\setlength{\tabcolsep}{4.5pt} 
\begin{tabular}{cc}
\hline
\begin{tabular}[c]{@{}c@{}}
Optimization
\end{tabular} &
Related STeP Features \\
\hline
    \arrayrulecolor{lightgray} 
\begin{tabular}[c]{@{}c@{}}
Dynamic\\
Tiling
\end{tabular} &
\begin{tabular}[c]{@{}c@{}}
Dynamic tile shape \\
Explicit memory hierarchy \\
\stepops{Accum} of dynamically-sized tiles\\
\end{tabular} \\ \hline
\begin{tabular}[c]{@{}c@{}}
Configuration \\
Time-multiplexing
\end{tabular} & 
\begin{tabular}[c]{@{}c@{}}
Explicit memory hierarchy \\
Dynamic Routing and Merging Operators \\
\end{tabular} \\ 
\hline
\begin{tabular}[c]{@{}c@{}}
Dynamic\\
Parallelization
\end{tabular} & Dynamic Routing and Merging Operators \\
    \arrayrulecolor{black}
\hline
\end{tabular}
\caption{Key STeP features that enable the optimizations}
\label{tab:optimizations}
\end{table}

\subsection{Dynamic Tiling} \label{subsec:dyn-tiling}

\begin{figure}
    \centering    \includegraphics[width=\columnwidth]{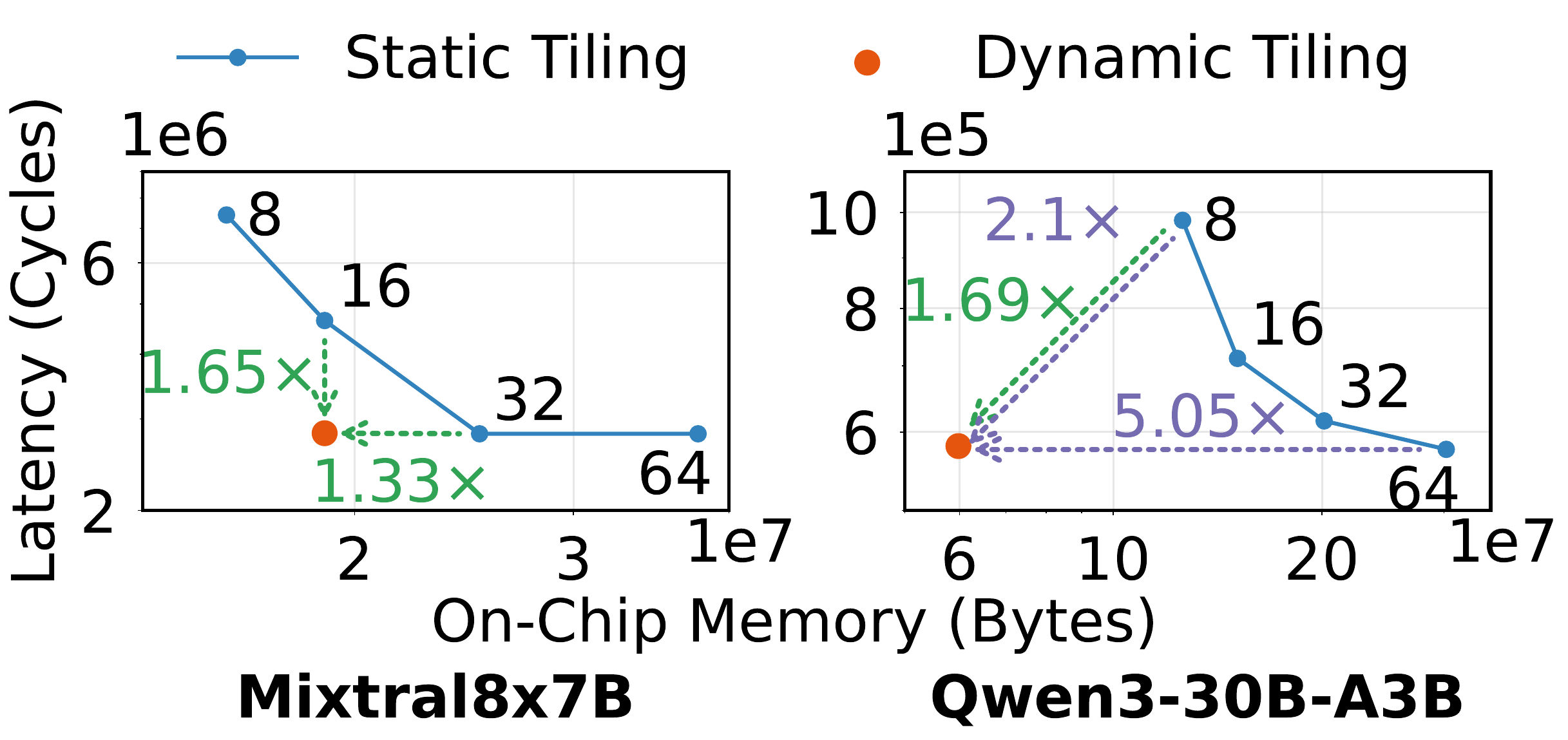}
    \caption{Performance and memory requirements of tiling strategies for the batch dimension of each expert (batch = 64). The numbers on the static tiling curve denote tile size.}
    \Description{A Pareto curve for Mixtral8x7B and Qwen3-30B-A3B with batch size of 64. The y-axis shows latency, and the x-axis shows on-chip memory usage.}
    \label{fig:dyn-tile-64-pareto}
\end{figure}

\begin{figure}
    \centering    \includegraphics[width=\columnwidth]{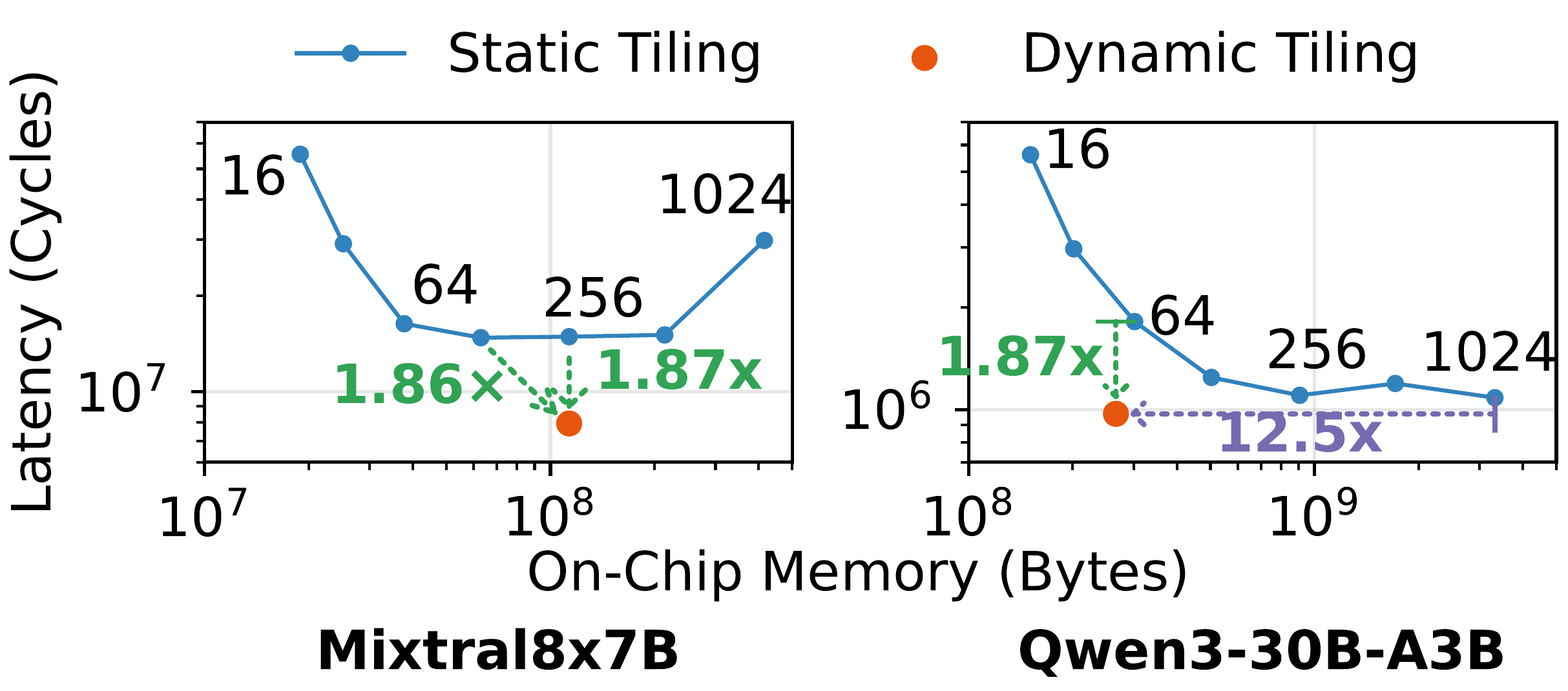}
    \caption{Performance and memory requirements of tiling strategies for the batch dimension (batch = 1024).}
    \Description{A Pareto chart comparing latency (cycles) and on-chip memory (bytes) for Mixtral-8x7B and Qwen3-30B-A3B by sweeping tile sizes from 16 to 1024. The plot shows that dynamic tiling extends the Pareto frontier beyond what is achievable with static tiling.}
    \label{fig:dyn-tile-1024-pareto}
    \Description{A Pareto curve for Mixtral8x7B and Qwen3-30B-A3B with batch size of 1024. The y-axis shows latency, and the x-axis shows on-chip memory usage.}
\end{figure}

Dynamic tiling is a scheduling strategy where the size of a tile in a data stream is determined at runtime.
When applying dynamic tiling to the batch dimension for MoEs, tokens are grouped into tiles whose size adapts to the number of tokens per expert in each batch. On the contrary, static tiling pads the tokens for each expert into statically-sized tiles.
Dynamic tiling can be expressed in STeP by replacing the first \stepops{Reshape} in \Cref{fig:big-example-step}
with a \stepops{Promote}. This enables the following \stepops{Accum} to accumulate dynamically-shaped tiles.

Dynamic tiling extends the Pareto frontier beyond what is achievable with static tiling. 
We demonstrate this using the performance improvements (green dashed arrows) and on-chip memory savings (purple dashed arrows) against the closest static-tiling Pareto point along each axis in \Cref{fig:dyn-tile-64-pareto,fig:dyn-tile-1024-pareto}.
As shown in \Cref{fig:dyn-tile-64-pareto}, for Mixtral8x7B, dynamic tiling achieves a $1.65\times$ speedup while using the same on-chip memory as $tile=16$. It also reduces on-chip memory by $1.33\times$ while delivering performance comparable to $tile=32$ (within $+0.26\%$ cycles).
Similarly, for Qwen3-30B-A3B, dynamic tiling achieves a $1.69\times$ speedup while using $2.1\times$ less on-chip memory than $tile=8$, and reduces on-chip memory by $5.05\times$ while maintaining performance comparable to $tile=64$ (within $+0.76\%$ cycles). The larger savings for Qwen3-30B-A3B are due to its higher number of experts.

At larger batch sizes, dynamic tiling enables design points that achieve performance unattainable using any static tile size, as shown in \Cref{fig:dyn-tile-1024-pareto}.
For Mixtral8x7B, static tiling saturates beyond $tile=128$, where increasing the tile size yields little to no additional speedup.
In contrast, dynamic tiling improves performance beyond $tile=128$, delivering a $1.86\times$ speedup while using $1.79\times$ more on-chip memory, and a $1.87\times$ speedup over $tile=256$ with the same on-chip memory requirement.
For Qwen3-30B-A3B, dynamic tiling achieves a $1.87\times$ speedup while using $1.13\times$ less on-chip memory than $tile=64$.
It also delivers a $1.12\times$ speedup while reducing on-chip memory by $12.5\times$ relative to the best-performing static configuration ($tile=1024$).

In summary, dynamic tiling breaks the prior Pareto-optimal frontier, achieving Pareto Improvement Distances (PID) of $1.33\times$ and $2.11\times$ for Mixtral-8$\times$7B and Qwen3-30B-A3B, respectively, in \Cref{fig:dyn-tile-64-pareto}, and $1.86\times$ and $1.87\times$ in \Cref{fig:dyn-tile-1024-pareto}.\footnote{For more details on PID, see \Cref{subsec:dyn-tiling-appendix}.} PID for a design point $p$ with respect to the Pareto-optimal subset of baseline points $F_B$ is defined as:
\[
\mathrm{PID}(p)
\;=\;
\min_{q \in F_B}\;
\max\!\left(
\frac{\mathrm{cycles}(q)}{\mathrm{cycles}(p)},
\frac{\mathrm{mem}(q)}{\mathrm{mem}(p)}
\right).
\]

The static-tiling Pareto curve reflects a trade-off between on-chip memory usage and off-chip traffic: small tiles incur frequent off-chip reloads, while large tiles waste on-chip memory due to padding and unused space.\footnote{The corresponding Pareto curve for off-chip traffic versus on-chip memory is shown in \Cref{subsec:dyn-tiling-appendix} of the Appendix.} Because the application is memory-bound under our simulated hardware configuration, this trade-off in off-chip traffic directly translates to a trade-off in performance. Dynamic tiling removes the need to choose between frequent reloads and wasted capacity by adapting tile sizes to the active workload. By achieving high performance with smaller or comparable on-chip memory, dynamic tiling frees up space that can be repurposed to further improve performance, for example by increasing data reuse via larger tiles or more aggressive operator fusion.

\subsection{Configuration Time-multiplexing}\label{subsec:config-timemultiplexing}
\begin{figure}
    \centering
    \includegraphics[width=0.9\columnwidth]{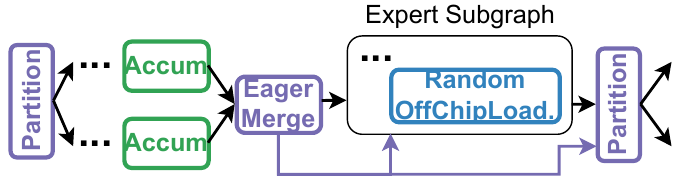}
    \caption{STeP graph with configuration time-multiplexing. Certain operators are omitted for simplicity.}
    \label{fig:config-time-multiplex-graph}
    \Description{The graph uses Partition - Reshape - Accum - EagerMerge and has a single expert region to express the time-multiplexing.}
\end{figure}

    Configuration time-multiplexing is an optimization that time-multiplexes a configuration across branches with shared computation structure in applications with data-dependent control flow. In the context of executing MoEs, a configuration is dynamically time-multiplexed across experts by routing inputs and weights accordingly. For the simplified MoE example in \Cref{fig:big-example-step}, configuration time-multiplexing can be expressed by inserting a pair of control-flow operators around the time-multiplexed region, as shown in \Cref{fig:config-time-multiplex-graph}. \stepops{EagerMerge} forwards inputs for each expert to the time-multiplexed region as soon as they become available. \stepops{RandomOffChipLoad} fetches the weight for the selected expert dynamically, instead of using \stepops{LinearOffChipLoad}.

\begin{figure}
    \centering    \includegraphics[width=\columnwidth]{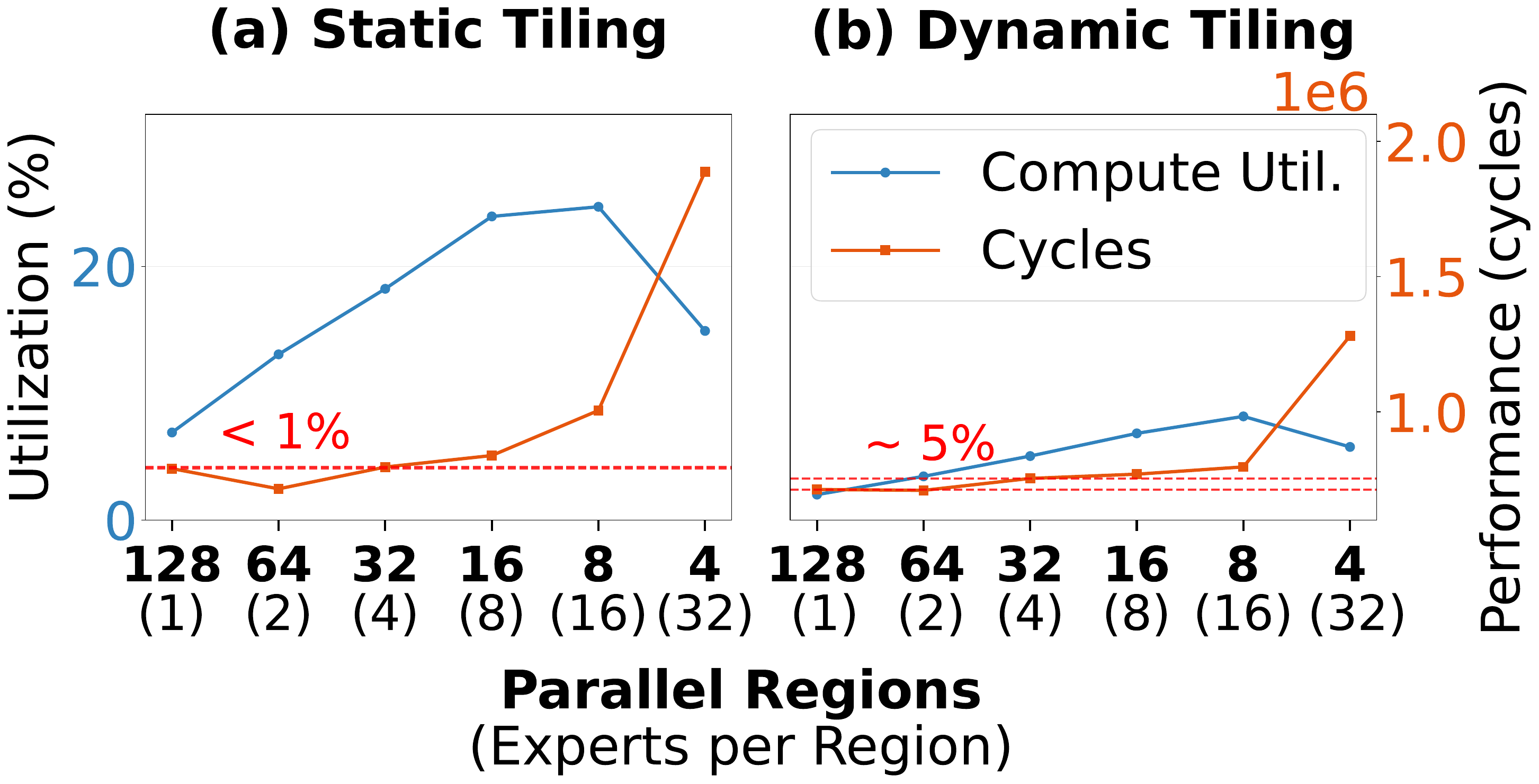}
    \Description{Line chart comparing compute utilization for the MoE layer in Qwen3-30B-A3B between static and dynamic tiling. Static tiling shows higher utilization but results in inflated FLOPs due to padding, while dynamic tiling achieves lower compute usage.}
\caption{Resource utilization for the MoE layer in Qwen3-30B-A3B with different tiling strategies for the batch dimension of each expert (batch size = 64). Dynamic tiling has lower compute utilization when compared to static tiling because static tiling has $3.81\times$ more total FLOPs due to padding. The tile size used in static tiling is 32.}
    \label{fig:time-multiplexing-tiling}
\end{figure}

\begin{figure}
    \centering    \includegraphics[width=0.9\columnwidth]{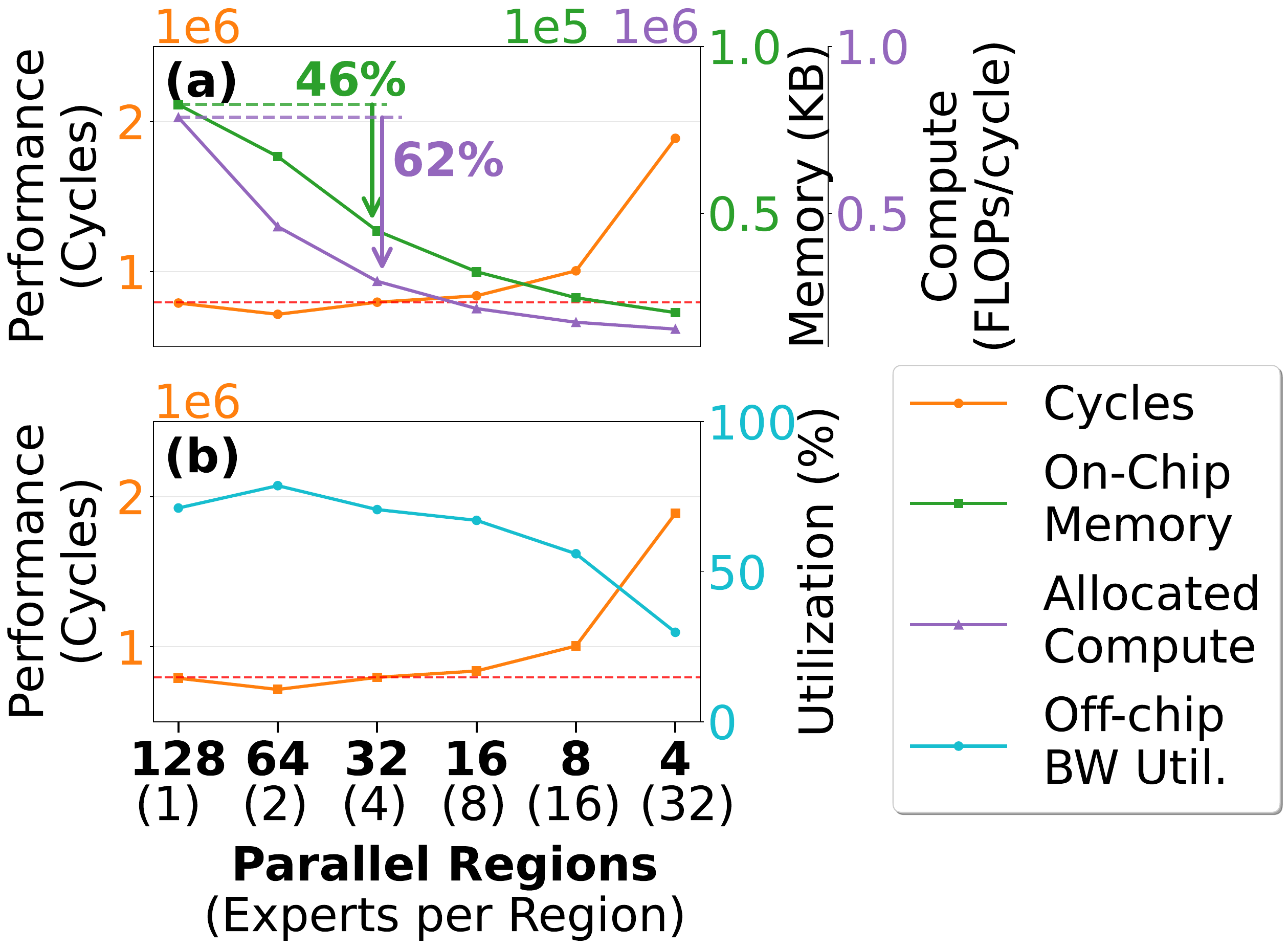}
    \caption{Resource usage and performance for the MoE layer in Qwen3-30B-A3B with time-multiplexing (tile size =32, batch size = 64). The compute utilization drop in \Cref{fig:time-multiplexing-tiling} is due to decreased off-chip memory bandwidth utilization because the number of parallel regions are not large enough to saturate the off-chip memory bandwidth.
}
    \label{fig:time-multiplexing-result}
    \Description{Two plots showing performance, memory usage, and compute throughput for the MoE layer in Qwen3-30B-A3B under varying degrees of time-multiplexing. As the number of parallel regions increases, performance and utilization trends are shown across cycles, memory, and bandwidth.}
\end{figure}

    Configuration time-multiplexing avoids allocating dedicated compute and memory resources for every possible branch.
    To quantify the resource savings, we sweep the number of experts sharing the same configured region for the MoE layer in Qwen3-30B-A3B using static tiling ($tile=32$) and dynamic tiling. With static tiling, compute utilization improves by $2.64\times$ with under 1\% performance overhead (\Cref{fig:time-multiplexing-tiling}(a)). With dynamic tiling, compute utilization improves by $2.51\times$ with about 5\% overhead (\Cref{fig:time-multiplexing-tiling}(b)).
    As shown in \Cref{fig:time-multiplexing-result}, this optimization achieves comparable performance while freeing up 62\% allocated on-chip compute and 46\% memory resources, which can be reallocated to support more concurrent requests or larger models.
    
\begin{figure*}[t]
    \centering
    \begin{minipage}[t]{0.33\textwidth}
        \centering
        \includegraphics[width=\textwidth]{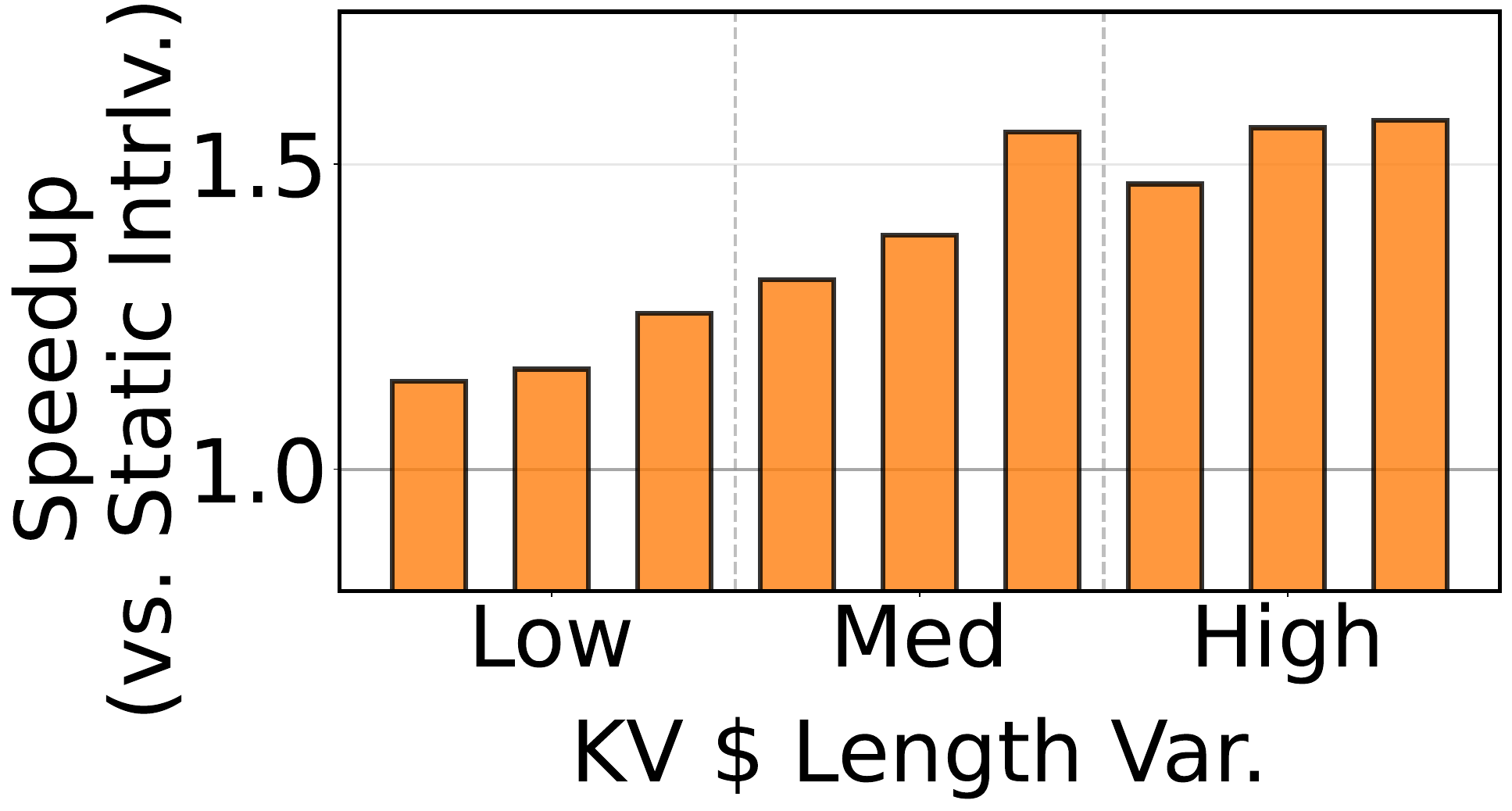}
        \caption{Speedup of dynamic parallelization over static interleaved parallelization across batches with different KV-cache length distributions (batch size=64).}
        \label{fig:interleave-dyn-comparison}
    \end{minipage}%
    \hspace{1em} 
        \begin{minipage}[t]{0.27\textwidth}
        \centering
        \includegraphics[width=\textwidth]{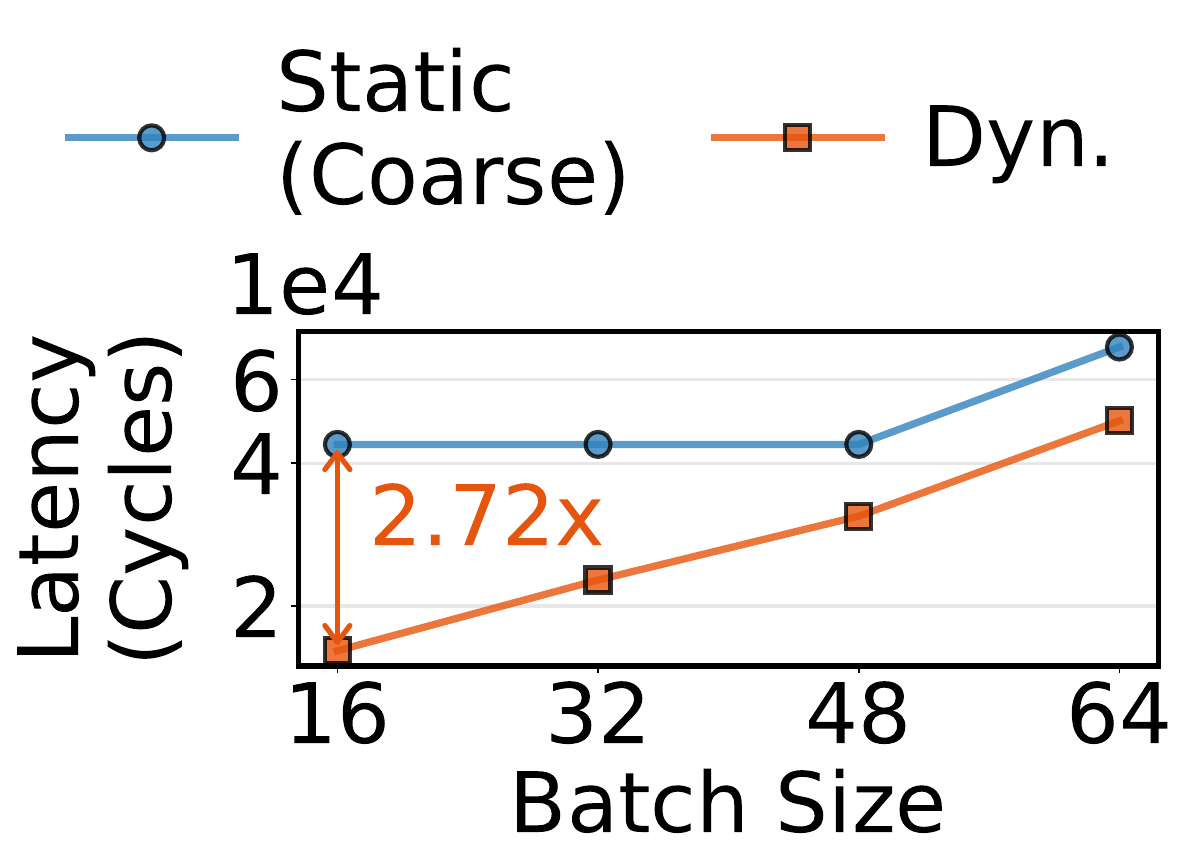}
    \caption{Performance comparison between coarse-grained and dynamic parallelization strategies.}
    \label{fig:coarse-dyn-comparison}
    \end{minipage}%
    \hspace{1em} 
    \begin{minipage}[t]{0.33\textwidth}
        \centering
        \includegraphics[width=\columnwidth]{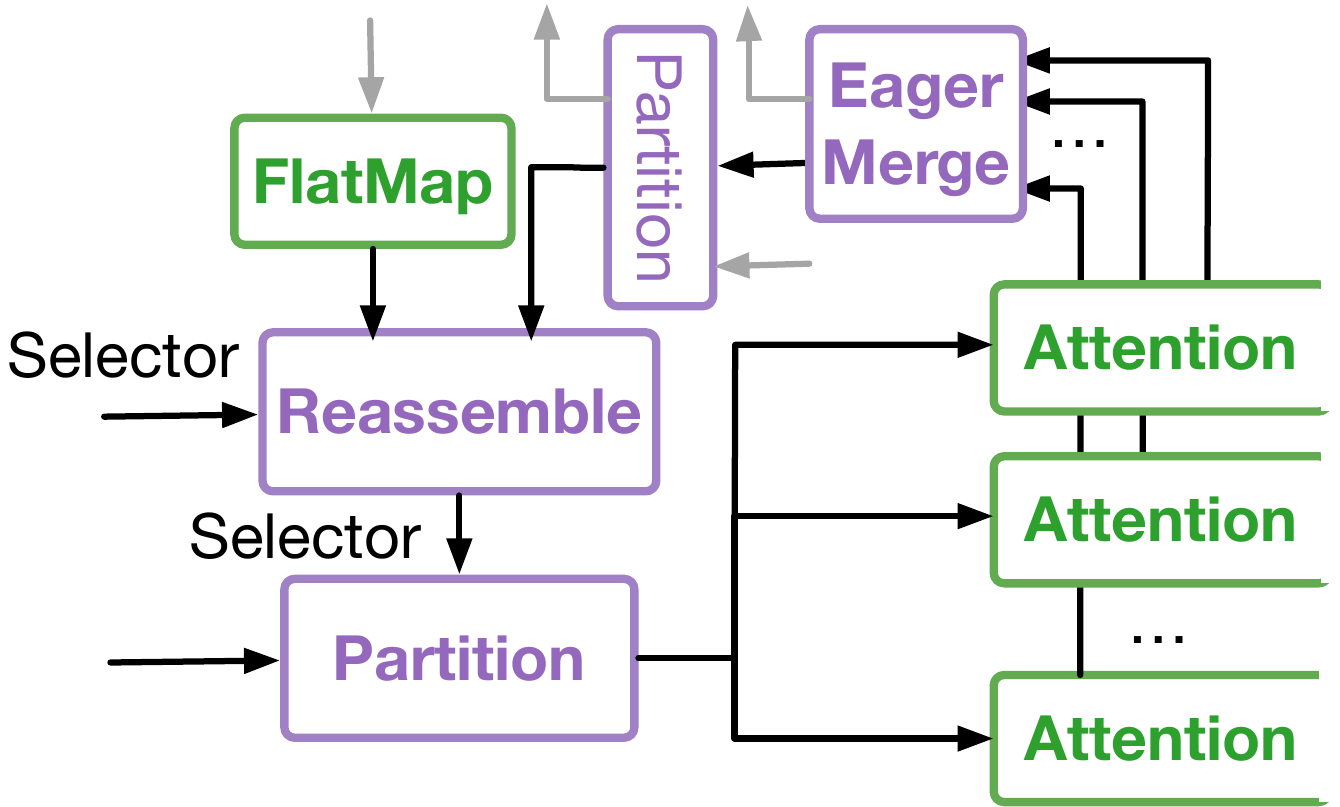}
        \caption{The STeP graph for dynamic parallelization. Shape operators omitted for simplicity.
        }
        \label{fig:dyn-par-graph}
    \end{minipage}%
    \Description{Three figures showing the performance improvement of dynamic parallelization and the corresponding STeP graph. In the STeP graph, the output of Partition is fed into each parallel attention region. The streams from each attention subgraph are fed into an EagerMerge. The output of EagerMerge and a FlatMap is fed into a Reassemble operator, which outputs the selector stream for the first Partition.}
\end{figure*}

\subsection{Dynamic Parallelization}\label{subsec:dyn-par}
    Dynamic parallelization is an optimization that dispatches work as soon as downstream parallel pipelines become available. This can improve performance by balancing load across spatially parallel regions when parallelizing workloads whose size or distribution can vary.
    In ML workloads, unevenly sized workloads arise in attention computation during decoding, since KV cache lengths vary across requests. The number of requests within a batch also varies dynamically due to optimizations such as continuous batching~\cite{yu2022orca,kwon2023pagedattention}.
    For attention, STeP implements dynamic parallelization as illustrated in \Cref{fig:dyn-par-graph}. Each request is routed to one of several parallel regions using \stepops{Partition}. The selector stream for \stepops{Partition} is formed by merging two streams: One for round-robin assignment of the initial iteration (\stepops{FlatMap}) and another signaling the availability of parallel regions (\stepops{EagerMerge}).

    We compare parallelization strategies across varying batch sizes and KV cache length distributions. We parallelize the batch dimension by four and use two tiled static parallelization baselines: coarse-grained and interleaved parallelization. Static coarse-grained parallelization fixes the number of requests assigned to each parallel region (16 in our implementation), whereas static interleaved parallelization distributes requests across regions in a round-robin fashion.
    
    As shown in \Cref{fig:interleave-dyn-comparison}, dynamic parallelization consistently outperforms static interleaved parallelization as KV cache length variation increases. With low variation, dynamic parallelization achieves $1.14\times$–$1.26\times$ speedup; with high variation, $1.47\times$–$1.57\times$ speedup. This is because under a larger KV cache length variation, static interleaved parallelization suffers from blocking when long requests increase load imbalance, leaving resources idle.
    
    Dynamic parallelization also maintains high utilization across parallel regions as batch size varies. In \Cref{fig:coarse-dyn-comparison}, it achieves a $2.72\times$ speedup over static coarse-grained parallelization at batch=16 because several parallel regions remain idle under static coarse-grained parallelization. Although static performance improves with larger batches, it remains $1.43\times$ slower at batch=64 due to persistent load imbalance.

\subsection{End-to-end Model}\label{subsec:end-to-end}
To evaluate end-to-end impact, we implement the full Qwen3-30B-A3B and Mixtral-8x7B models in STeP, with and without the proposed optimizations presented above. Each Transformer decoder layer is fused into a single STeP graph and executed repeatedly with layer-specific weights. Each layer comprises of STeP graphs for QKV generation, attention, and MoE; we parallelize the batch dimension by a factor of four for QKV generation and attention, and use expert parallelism for MoE.
We use static interleaved parallelization for attention, which performs best overall across batch sizes and KV-cache lengths in our ablation study (\Cref{fig:dyn-par-result} in the Appendix).
For MoE computation, we use static tiling with performance- and memory-matched tile sizes, which are the same closest points along each axis, from \Cref{fig:dyn-tile-64-pareto}.

\begin{figure}[t]
    \centering  
    \includegraphics[width=\linewidth]{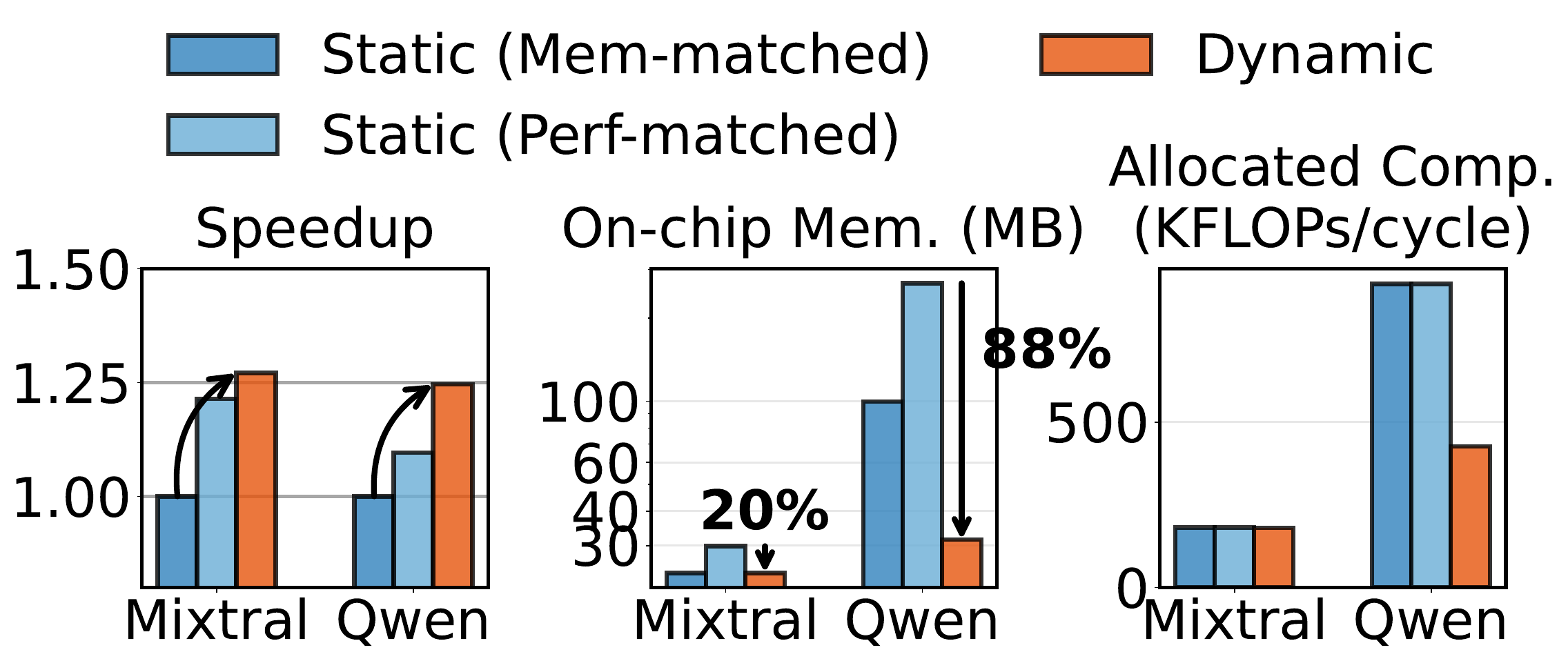}
    \caption{End-to-end result for Qwen3-30B-A3B and Mixtral8x7B. We use KV-cache length traces with median standard deviation and expert routing traces whose expert bin count standard deviation matches the overall average.}
    \Description{Three grouped bar plots each showing the speedup, on-chip memory usage, and allocated compute resources for the memory-matched static implementation, the performance-matched static implementation, and the dynamic implementation.}
    \label{fig:end-to-end}
\end{figure}

As shown in \Cref{fig:end-to-end}, our proposed optimizations improve performance while using comparable or fewer on-chip resources.
Mixtral-8×7B and Qwen3-30B-A3B achieve $1.27\times$ and $1.15\times$ speedups, respectively, over the memory-matched implementation, driven by fewer off-chip accesses from dynamic tiling and better load balancing from dynamic parallelization.
Qwen3-30B-A3B achieves this speedup while using $69\%$ less on-chip memory and $54\%$ fewer compute resources through configuration time-multiplexing.

Compared with the performance-matched static implementation, on-chip memory usage is reduced by 20\% for Mixtral-8×7B and by 88\% for Qwen3-30B-A3B due to dynamic tiling, with additional savings for Qwen3-30B-A3B from configuration time-multiplexing. Even with the static performance-matched MoE implementation in \Cref{fig:dyn-tile-64-pareto}, the dynamic implementation still deliver $1.05\times$ (Mixtral-8$\times$7B) and $1.14\times$ (Qwen3-30B-A3B) speedups due to dynamic parallelization.
We do not apply configuration time-multiplexing to Mixtral-8×7B, since all experts are active at a batch=64. However, many recent MoE models activate only a small fraction of a large expert pool (128+ experts) per token~\cite{mixtral,meta2025llama4,openai2025gptoss120bgptoss20bmodel,kimiK2}, indicating that the resource savings observed for Qwen3-30B-A3B generalize to modern MoEs.

\subsection{DSE with the Symbolic Frontend and Simulator}\label{subsec:dse}

The experiments in the preceding subsections not only demonstrate new optimizations, but also illustrate how STeP’s symbolic frontend and simulator can be used for design space exploration (DSE).
In the dynamic tiling experiments (\Cref{fig:dyn-tile-64-pareto,fig:dyn-tile-1024-pareto}), static tiling requires sweeping tile sizes to identify the optimal trade-off between on-chip memory and performance. If the hardware supports only static tiling, the symbolic frontend and simulator can be used to search for the optimal tile size for a given model, batch size, and expert distribution.
Similarly, for configuration time-multiplexing (\Cref{fig:time-multiplexing-result,fig:time-multiplexing-tiling}) and dynamic parallelization (\Cref{fig:coarse-dyn-comparison,fig:interleave-dyn-comparison}), the symbolic frontend and simulator enable comparison of design points across different schedules, such as varying degrees of time-multiplexing and parallelization strategies. They also allow evaluation under input variations, including batch size and KV-cache length distributions.

\section{Discussion on Future Work}\label{sec:discussion}
This section discusses future support for compiling models defined in a high-level framework to STeP, and potential approaches to support the dynamic features of STeP on SDAs. 
We leave an optimal hardware SDA design and a high-level compiler for STeP as future work. 

\subsection{Compiling from High-level Frameworks to STeP}
\label{subsec:compile}
Although we present STeP as a programming abstraction in this paper, it can also serve as an intermediate representation for compilers. For example, ML models defined in high-level frameworks such as PyTorch can be compiled to STeP using the \texttt{torch.compile} interface~\cite{ansel2024pytorch}. 
\texttt{torch.compile} captures the model into an FX graph expressed in terms of tensor-level operators, which a compiler can traverse and systematically lower into corresponding STeP subgraphs. 
Optimization schedules—including parallelization, tiling, and configuration time-multiplexing—can be specified over the index variables of the program. 
These schedules guide how each FX node is translated into STeP and determine which optimizing rewrites are applied during lowering.

\subsection{Supporting Dynamic STeP features in SDAs}\label{subsec:dynamic-hardware}
Prior SDAs have already demonstrated architectures that process stop-token-embedded data streams \cite{onyx, revet}. A few bits in the datapath are used to identify the stop tokens and their level, and the streams are processed either by repurposing existing hardware units ~\cite{revet} or by designing a new dedicated state machine to process stop tokens~\cite{onyx}.

STeP’s control flow operators (\Cref{subsubsec:route-merge}) can be implemented by spatially laying out all branches and activating the appropriate ones data-dependently at runtime. Routing can be implemented either through predication in compute units~\cite{sara} or within the network-on-chip interconnect~\cite{riptide}.

To handle dynamic tensor sizes, the memory system must support virtualization by allocating space at a fixed granularity independent of stream length and maintaining mappings between stream references and their memory addresses. Noncontiguous allocation is also required to avoid fragmentation. This can be implemented using a hardware-managed mapping cache (e.g. a linked list) that translates stream references into a sequence of noncontiguous physical addresses. With 512~KB of local memory per unit~\cite{sn40l}, the mapping cache requires less than 30~KB of metadata ($\approx$6\% overhead), comparable to the tag overhead in conventional caches.
Furthermore, arbitrary tensor sizes without an upper bound can be supported via spilling mechanisms demonstrated in prior SDAs~\cite{ripple}, where data is automatically spilled and metadata for accessing the spilled data remains on-chip.

\section{Related Work}\label{sec:related-work}
This section presents work related to STeP beyond SDAs. 

\textbf{CUDA graphs and conditional nodes}~\cite{nvidia_cuda_programming_guide} enable conditional or repeated execution of subgraphs without returning control to the CPU. While similar to STeP’s control-flow operators, they do not support dynamic shapes.
Common workarounds map dynamic shapes to an enumeration of many static GPU kernels, which adds overhead as the dynamic dimension's range increases~\cite{pytorch_cudagraph_trees}.

\textbf{Dynamic task-parallelism frameworks}~\cite{taskstream,taskflow,cheng2023beyond} share several themes with STeP but operate at different granularities and use different mechanisms. 
TaskStream~\cite{taskstream} and STeP both support asynchronous units, dynamic work distribution, and dynamic data reuse. However, STeP realizes these ideas at the tile granularity via dynamic dataflow blocks, whereas TaskStream targets coarser task instances via hardware scheduling. Taskflow~\cite{taskflow} also provides in-graph control flow and graph-based parallel abstractions, but targets CPU/GPU systems and supports dynamic parallelism via work stealing. Cheng et al.~\cite{cheng2023beyond} similarly target irregular workloads on scratchpad-managed architectures, but focus on a Cilk/TBB-style work-stealing runtime for manycore systems.

\section{Conclusion} 

We introduced the \pluralstep, a streaming abstraction for dynamic tensor applications on spatial dataflow accelerators. 
STeP expresses optimizations that are Pareto-optimal over prior spatial dataflow abstractions, delivering speedups and/or with fewer resources.
Its stream structure and shape semantics expose performance-critical metrics, creating new opportunities for optimization. We envision that STeP will enable richer forms of dynamic applications and architectures.

\section{Acknowledgments}
We thank Paul Mure, Rubens Lacouture, Christophe Gyurgyik, Suguna Velury, Tanmay Garg, Benjamin Driscoll, Raghu Prabhakar, Alex Rucker, Fredrik Kjolstad, Tian Zhao, Shiv Sundram, Qizheng Zhang, and Sally Wang for discussion and their helpful feedback. 
Gina Sohn was supported by the Stanford Graduate Fellowship and NSF GRFP. 
This work was supported in part by DARPA under the Machine learning and Optimization-guided Compilers for Heterogeneous Architectures (MOCHA) program (award number HR00112520038), and by the Naval Surface Warfare Center under Agreement
No. N00164-23-9-G057-01. This research was also supported in part by the Stanford Data Analytics for What’s Next (DAWN) Affiliate Program.
Any opinions, findings, and conclusions
or recommendations expressed in this material are those of
the authors and do not necessarily reflect the views of the
aforementioned funding agencies.


\bibliographystyle{ACM-Reference-Format}
\balance
\bibliography{references}

\newpage

\appendix

\section{Artifact Appendix}
\subsection{Abstract}
This appendix describes how to set up and run programs written in the Symbolic Python frontend for the Streaming Tensor Program (STeP) using the STeP Rust simulator and the Bluespec SystemVerilog HDL simulator. The artifact provides a Docker image containing all required dependencies (Python, Rust, Bluespec, protobuf, etc.) and scripts to reproduce the experimental results reported in the paper. The artifact can be executed on any x86-64 machine with Docker, Python 3, Git, and Bash support, at least \minmem\ GB of RAM, and more than 20 GB of disk space.

\subsection{Artifact Check-List (Meta-Information)}\label{subsec:artifact-info}
\begin{itemize}
    \item \textbf{Data set:} AzureLLMInferenceDataset~\cite{splitwise}, expert routing data collected by running Qwen3-30B-A3B~\cite{qwen3} and Mixtral8x7B~\cite{mixtral} models using the \texttt{hh-rlhf} serving trace~\cite{hh-rlhf}, and synthetically generated tensors. To reduce artifact size, we include only the data used to generate the plots in the paper from the AzureLLMInferenceDataset and the expert routing data.
    \item \textbf{Run-time environment:} Docker, Git, Python~3, and Bash must be installed on the local machine. Proficiency in Bash and Git is recommended.
    \item \textbf{Hardware:} Any conventional x86-64 CPU with at least \minmem\ GB of RAM should work.
    \item \textbf{Metrics:} Cycles, off-chip memory traffic, on-chip memory requirements, allocated compute resources, compute resource utilization, and off-chip memory bandwidth utilization.
    \item \textbf{Output:} Terminal output, files, and graphs (PDF figures).
    \item \textbf{How much disk space is required (approximately):} Approximately 20~GB of disk space is sufficient.
    \item \textbf{How much time is needed to prepare the workflow (approximately)?:} About 10--15 minutes.
    \item \textbf{How much time is needed to complete experiments (approximately)?:}
    The total time to complete all experiments is approximately 24.5 hours when measured on a Google Cloud C4-standard-8 instance (8 Intel Emerald Rapids vCPUs, 30~GB memory). The breakdown is as follows:
    \begin{itemize}
        \item \Cref{fig:step-vs-bluespec}: 2 hours
        \item \Cref{fig:dyn-tile-64-pareto}: 2 hours 30 minutes
        \item \Cref{fig:dyn-tile-1024-pareto}: 17 hours 10 minutes
        \item \Cref{fig:time-multiplexing-tiling}: 1 hour 40 minutes
        \item \Cref{fig:time-multiplexing-result}: 50 minutes
        \item \Cref{fig:interleave-dyn-comparison}: 4 minute
        \item \Cref{fig:coarse-dyn-comparison}: 1 minutes
        \item \Cref{fig:dyn-par-result}: 15 minutes
    \end{itemize}
    \item \textbf{Publicly available?:} Yes, on GitHub at \href{https://github.com/stanford-ppl/step_artifact}{\color{blue}step\_artifact} and \href{https://github.com/stanford-ppl/step-artifact-hdl}{\color{blue}step-artifact-hdl}.
    \item \textbf{Code licenses (if publicly available)?:} MIT License
    \item \textbf{Workflow framework used?:} Docker
    \item \textbf{Archived (provide DOI)?:} Yes. The DOI is \newline
    \href{https://doi.org/10.6084/m9.figshare.31095274}{\color{blue}https://doi.org/10.6084/m9.figshare.31095274}
\end{itemize}

\subsection{Description}

\subsubsection{How to Access}
The code for this submission can be downloaded from the \href{https://github.com/stanford-ppl/step_artifact}{\color{blue}step\_artifact} and \href{https://github.com/stanford-ppl/step-artifact-hdl}{\color{blue}step-artifact-hdl} repositories. The \href{https://github.com/stanford-ppl/step_artifact}{\color{blue}step\_artifact} repository includes a Dockerfile that can be used to build the Docker image for the full evaluation of the artifact. The Docker image is also available at \href{https://doi.org/10.6084/m9.figshare.31095274}{\color{blue}https://doi.org/10.6084/m9.figshare.31095274}.

\subsubsection{Hardware Dependencies}
We recommend using an x86-64 machine with at least \minmem\ GB of memory.

\subsubsection{Software Dependencies}
The artifact requires a machine with Docker, Git, and Python~3 installed. We evaluated the artifact using the following configuration: Ubuntu 24.04 LTS, Docker 29.1.3, and Python 3.12 (Intel-based machine).

\subsubsection{Data Sets}
The experiments for \Cref{fig:dyn-tile-64-pareto,fig:dyn-tile-1024-pareto,fig:time-multiplexing-tiling,fig:time-multiplexing-result} use expert routing data collected by running the Qwen3-30B-A3B~\cite{qwen3} and Mixtral8x7B~\cite{mixtral} models with the \texttt{hh-rlhf} serving trace~\cite{hh-rlhf}. To select representative cases, we measure the standard deviation of expert bin counts across iterations and layers and choose the case whose deviation is closest to the overall average.

The experiment for \Cref{fig:dyn-par-result,fig:coarse-dyn-comparison,fig:interleave-dyn-comparison} uses KV cache lengths sampled from the AzureLLMInferenceDataset~\cite{splitwise}. We analyze 5{,}000 requests within a time window, forming batches with varying prompt-length distributions. We experiment with batches whose prompt-length standard deviation matches that of the full set, as well as batches with the most and least variability.
The KV cache length data and expert routing data used for the experiments are included in the \href{https://github.com/stanford-ppl/step_artifact}{step\_artifact} repository.

\subsection{Installation}
To install the artifact, first clone the \href{https://github.com/stanford-ppl/step_artifact}{\color{blue}step\_artifact} and \href{https://github.com/stanford-ppl/step-artifact-hdl}{\color{blue}step-artifact-hdl} repositories to the local machine. Then, build the Docker image using the following commands (the build can take up to 5 minutes):

\begin{lstlisting}[
  basicstyle=\ttfamily\small,
  breaklines=true,
  language=bash,
  keywordstyle=\color{black},
  commentstyle=\color{black},
  stringstyle=\color{black},
  identifierstyle=\color{black},
  emphstyle=\color{black},
  literate={}]
$ git clone --recursive https://github.com/stanford-ppl/step_artifact.git
$ git clone https://github.com/stanford-ppl/step-artifact-hdl.git
$ docker build -f step_artifact/Dockerfile 
    -t step_artifact .
\end{lstlisting}

The Docker container can be started with the following command, which will print the container ID:
{\small
\begin{verbatim}
$ docker run -dit step_artifact bash
\end{verbatim}
}

The container can be attached by running:
{\small
\begin{verbatim}
$ docker attach <CONTAINER_ID>
\end{verbatim}
}

Once inside the Docker container, move into the directory \texttt{step\_artifact} and run the following command to set up the environment:

\begin{lstlisting}[
    language=bash,
    basicstyle=\ttfamily\small,
    breaklines=true,
    commentstyle=\color{green!60!black}
]
### Inside the Docker container ###
$ cd /root/step_artifact
$ source setup.sh
\end{lstlisting}

\subsection{Experimental Workflow}
The experimental workflow for this artifact consists of running scripts inside the Docker container to execute experiments and generate the figures in the paper. Detailed instructions can be found in the \texttt{README.md} files within the \href{https://github.com/stanford-ppl/step_artifact}{\color{blue}step\_artifact} and \href{https://github.com/stanford-ppl/step-artifact-hdl}{\color{blue}step-artifact-hdl} repositories.

\subsection{Evaluation and Expected Results}

All experiments and figures can be reproduced using the following commands. In total, the workflow takes approximately 7 hours when tested on a Google Cloud C4-standard-8 instance (8 Intel Emerald Rapids vCPUs, 30~GB memory).

\begin{lstlisting}[
    language=bash,
    basicstyle=\ttfamily\small,
    breaklines=true,
    commentstyle=\color{green!60!black},
    deletekeywords={eval},
    literate={}  
]
### In the Docker container ###
$ cd /root/step_artifact
# Figures 9, 10, 12, 13, 14, 15, and 21
$ source ae_cmd.sh
# Figure 8
$ cp /root/step_artifact/hdl_validation/fig8.csv /root/step-artifact-hdl/step_reference.csv
$ cd /root/step-artifact-hdl
$ ./run_dse_and_figure.sh
# ctrl+p ctrl+q
\end{lstlisting}

Once the experiments finish, detach from the container by pressing \texttt{Ctrl+P} followed by \texttt{Ctrl+Q}. To copy the experiment results and figures from the container, move into the cloned \texttt{step\_artifact} repository on the local machine and run the following commands. The \texttt{CONTAINER\_ID} is the same ID used to attach to the container; it can also be retrieved by running \texttt{docker ps}.
The results and figures will be copied to \code{step_artifact/results}.

\begin{lstlisting}[
    language=bash,
    basicstyle=\ttfamily\small,
    breaklines=true,
    commentstyle=\color{green!60!black}
]
### On the local machine ###
$ cd step_artifact
$ mkdir -p results
$ python copy_from_docker.py --docker_id 
    <CONTAINER_ID> --output_dir ./results
\end{lstlisting}

The expected directory structure under \code{step_artifact/results} is as follows:
{\small
\begin{verbatim}
step_artifact/results
|_ step-artifact-hdl
|_ step_artifact
    |_ dyn_tiling
    |_ dynamic_par
    |_ timeshare_mem_bound
\end{verbatim}
}

\begin{itemize}
    \item \Cref{fig:step-vs-bluespec}: The reproduced figure and experiment results can be found in the \texttt{step-artifact-hdl} directory. The file \code{validation.pdf} should match \Cref{fig:step-vs-bluespec}. The values used to generate the plot are provided in the other two CSV files in the same directory.
    
    \item \Cref{fig:dyn-tile-64-pareto}: The reproduced figure and experiment results can be found in the \texttt{dyn\_tiling} directory. The file \code{figure9.pdf} should match \Cref{fig:dyn-tile-64-pareto}. The values used to generate the plot can be found in \code{figure_9_mixtral_b64_raw.csv} and \code{figure_9_qwen_b64_raw.csv}.
    
    \item \Cref{fig:dyn-tile-1024-pareto}: The reproduced figure and experiment results can be found in the \texttt{dyn\_tiling} directory. The file \code{figure10.pdf} should match \Cref{fig:dyn-tile-1024-pareto}. The values used to generate the plot can be found in \code{figure_10_mixtral_b1024_raw.csv} and \code{figure_10_qwen_b1024_raw.csv}.
    
    \item \Cref{fig:time-multiplexing-tiling}: The reproduced figure and experiment results can be found in the \code{timeshare_mem_bound} directory. The file \code{figure12.pdf} should match \Cref{fig:time-multiplexing-tiling}. The values used to generate the plot are provided in the remaining CSV files in the same directory.
    
    \item \Cref{fig:time-multiplexing-result}: The reproduced figure and experiment results can be found in the \code{timeshare_mem_bound} directory. The file \code{figure13.pdf} should match \Cref{fig:time-multiplexing-result}. 
    The values used to generate the plot are provided in the remaining CSV files in the same directory. 

    \item \Cref{fig:interleave-dyn-comparison}: The reproduced figure and experiment results can be found in the \code{dynamic_par} directory. The file \code{figure14.pdf} should match \Cref{fig:interleave-dyn-comparison}. The values used to generate the plot are provided in \code{batch64_interleave_dynamic.csv}.
    
    \item \Cref{fig:coarse-dyn-comparison}: The reproduced figure and experiment results can be found in the \code{dynamic_par} directory. The file \code{figure15.pdf} should match \Cref{fig:coarse-dyn-comparison}. The values used to generate the plot are provided in \code{batch_sweep_coarse_vs_dynamic.csv}.

    \item \Cref{fig:dyn-par-result}: The reproduced figure and experiment results can be found in the \code{dynamic_par} directory. The file \code{figure21.pdf} should match \Cref{fig:dyn-par-result}. The values used to generate the plot are provided in the remaining CSV files in the same directory.
\end{itemize}

\subsection{Toolchain Customization}
Details on how to customize the toolchain (the Python frontend and Rust simulator) can be found in the \href{https://github.com/stanford-ppl/step_artifact?tab=readme-ov-file#optional-to-customise-or-extend-the-toolchain}{\color{blue}{To customize or extend the toolchain}} section of the \texttt{README.md} file in the {\color{blue}step\_artifact} repository.


\section{Appendix}
\subsection{STeP Operator Syntax and Shape Semantics}
This section contains the syntax and shape semantics of STeP operators. We express stream types in the form of \texttt{Strm<T,a>} where \texttt{T} is the data type of the stream and \texttt{a} is the rank of the stream. We will use uppercase letters in the angle brackets (\texttt{<,>}) to denote the data type of the stream and lowercase letters for the stream rank.

We use different uppercase letters to express the available data types for each operator.
\begin{itemize}
    \item \texttt{R,R$'$}: Any data type
    \item \texttt{A,B}: Non-buffer type
    \item \texttt{S}: Statically sized tile
    \item \texttt{SEL}: Selector type
    \item \texttt{I}: $[1,1]$ tile of integer address data type.
\end{itemize}

For dynamic routing and merging operators (\Cref{tab:step-route-merge}), the subscript $i$ in the input and output stream shape is used to specify the shape of the $i$-th input or output stream. For the Reshape operator, when splitting a dimension higher than the innermost (scalar) dimension, it should be a static dimension divisible by the chunk size. When splitting the innermost (scalar) dimension, there is no restriction on the dimension shape, and it will be accordingly padded.
 
\begin{table*}
    \centering
    \small 
    \setlength{\tabcolsep}{3pt} 
    \begin{tabular}{llll}
    \hline
    \multicolumn{1}{c}{Operator Signature} & In Stream Shape & Out Stream Shape &  \\ \hline
    \arrayrulecolor{lightgray}
    \texttt{\tableop{LinearOffChipLoad}<S,R,a,b> (ref: Strm<R,b>, base\_addr: int, } & $[D_b,\cdots,D_0]$  & $[D_b,\cdots,D_0,$ &  \\
    \phantom{\texttt{\tableop{LinearOffChipLoad}<S,R,a,b> (}}\texttt{tiled\_in\_shape: [int],}&  & \ \ $D'_{a-1},\cdots,D'_0]$ &  \\
    \phantom{\texttt{\tableop{LinearOffChipLoad}<S,R,a,b> (}}\texttt{stride: [int], tiled\_out\_shape: [int])}&  & (\texttt{a=|tiled\_in\_shape|}) & \\
    \phantom{\texttt{\tableop{LinearOffChipLoad}<S,R,a,b> (}}\texttt{$\rightarrow$ Strm<S,a+b>}&  &  & \\ \cline{1-3}
    
     \texttt{\tableop{LinearOffChipStore}<S,a> (in: Strm<S,a>, base\_addr: int)} & $[D_{a-1},\cdots,D_0]$ &  & \\ \cline{1-3}
    
     \texttt{\tableop{RandomOffChipLoad}<I,S,a,b> (raddr: Strm<I,a>, base\_addr: int, } &  $[D_a,\cdots,D_0]$ & $[D_a,\cdots,D_0]$  & \\  
     \phantom{\texttt{\tableop{RandomOffChipLoad}<I,S,a,b> (}}\texttt{tiled\_in\_shape: [int]) $\rightarrow$ Stream<S,a>} &   & & \\ \cline{1-3}

     \texttt{\tableop{RandomOffChipStore}<I,S,a,b> (waddr: Strm<I,b>, wdata: Strm<S,b>,} &  $[D_a,\cdots,D_0] \ \texttt{(waddr)}$ & $[D_a,\cdots,D_0]$  & \\  
     
     \phantom{\texttt{\tableop{RandomOffChipStore}<I,S,a,b> (}}\texttt{base\_addr: int, tiled\_in\_shape: [int]) } & $[D'_b,\cdots,D'_0]\ \texttt{(wdata)}$  & & \\ 
     
     \phantom{\texttt{\tableop{RandomOffChipStore}<I,S,a,b> (}}\texttt{$\rightarrow$ Stream<bool,a>} &  & & \\ 
     \arrayrulecolor{black}\hline
    
    \end{tabular}
    \caption{STeP off-chip memory operators. The square brackets in the operator signature express a list type.
}
    \label{tab:step-off-chip}
\end{table*}

\begin{table*}
    \centering
    \small 
    \setlength{\tabcolsep}{3pt} 
    \begin{tabular}{llll}
    \hline
    \multicolumn{1}{c}{Operator Signature} & In Stream Shape & Out Stream Shape &  \\ \hline

    \arrayrulecolor{lightgray}
     \texttt{\tableop{Bufferize}<S,a,b> (in: Strm<S,a>) $\rightarrow$ Strm<Buffer<S,b>,a-b>} & $[D_a,\cdots,D_b,$ & $[D_a,\cdots,D_b]$ & \\ 
     & \ \ $D_{b-1},\cdots,D_0]$ & (buffer: $[D_{b-1},\cdots,D_0]$) & \\ \cline{1-3}

     \texttt{\tableop{Streamify}<S,R,a,b,c> (in: Strm<Buffer<S,a>,b>, ref: Strm<R,b+c>,} & $[D_b,\cdots,D_0]\ {(\texttt{data})}$ & $[D_b,\cdots,D_0,D'_{c-1},\cdots,D'_0,$ & \\ 
     
     \phantom{\texttt{\tableop{Streamify}<S,R,a,b,c> (}}\texttt{stride: [int], out\_shape: [int])} & $[D_b,\cdots,D_0,$  & \ \ $D''_{|out\_shape|-1},\cdots,D''_0]$ & \\ 

     \phantom{\texttt{\tableop{Streamify}<S,R,a,b,c> (}}\texttt{$\rightarrow$ Strm<S,|out\_shape|+b+c>} & \ \ $D'_{c-1},\cdots,D'_0]\ \texttt{(ref)}$ &  & \\ 

     \arrayrulecolor{black}\hline

    \end{tabular}
    \caption{STeP on-chip memory operators. For Streamify, if the buffer is dynamically-sized, \texttt{|out\_shape|} is replaced with \texttt{a}.
    }
    \label{tab:step-on-chip}
\end{table*}

\begin{table*}
    \centering
    \small 
    \setlength{\tabcolsep}{5pt} 
    \begin{tabular}{llll}
    \hline
    \multicolumn{1}{c}{Operator Signature} & In Stream Shape & Out Stream Shape &  \\ \hline
    \arrayrulecolor{lightgray}

     \texttt{\tableop{Map}<A,B,a> (in: Strm<A,a>, fn: Fn(A)$\rightarrow$ B) $\rightarrow$ Strm<B,a>}  &  $[D_a,\cdots,D_0]$ & $[D_a,\cdots,D_0]$ & \\ \cline{1-3}
    
     \texttt{\tableop{Accum}<A,R,a,b> (in: Strm<A,a>, update\_fn: Fn(A,R)$\rightarrow$R,} & $[D_a,\cdots,D_b,$ & $[D_a,\cdots,D_b]$ & \\ 
     
     \phantom{\texttt{Accum<A,R,a> (}}\texttt{\ init\_fn: Fn()$\rightarrow$R) $\rightarrow$ Strm<R,a-b>} & \ \ $D_{b-1},\cdots,D_0]$ &  & \\  \cline{1-3}
    
     \texttt{\tableop{Scan}<A,B,a,b> (in: Strm<A,a>, update\_fn: Fn(A,B) $\rightarrow$ B, } & $[D_a,\cdots,D_{b},$ & $[D_a,\cdots,D_{b},$ & \\ 

     \phantom{\texttt{Scan<A,B,a,b> (}}\texttt{init\_fn: Fn() $\rightarrow$ B) $\rightarrow$ Strm<B,a>} & \ \ $D_{b-1},\cdots,D_0]$ & \ \ $D_{b-1},\cdots,D_0]$ & \\ \cline{1-3}
    
     \texttt{\tableop{FlatMap}<A,B,a,b> (in: Strm<A,a>, fn: Fn(A)$\rightarrow$Strm<B,b>) $\rightarrow$ Strm<B,a+b>} & $[D_a,\cdots,D_1,D_0]$ & $[D_a,\cdots,D_1,D'_b,\cdots,D'_0]$ & \\ \arrayrulecolor{black} \hline

    \end{tabular}
    \caption{STeP higher-order operators.
    }
    \label{tab:step-higher-order}
\end{table*}

\begin{table*}
    \centering
    \small 
    \setlength{\tabcolsep}{3pt} 
    \begin{tabular}{llll}
    \hline
    \multicolumn{1}{c}{Operator Signature} & In Stream Shape & Out Stream Shape &  \\ \hline
    \arrayrulecolor{lightgray}
     \texttt{\tableop{Partition}<R,SEL,a,b> (in: Strm<R,a>, sel: Strm<SEL,b>,} & $[D_a,\cdots,D_0]\ \texttt{(in)}$ & $[D^i_{a-b},D^i_{a-b-1},\cdots,D^i_0]_i$ & \\ 
     
     \phantom{\texttt{\tableop{Partition}<R,SEL,a,b>(}} \texttt{num\_consumers: int) $\rightarrow$ [Strm<R,a-b>]} & $[D_a,\cdots,D_{a-b}]\ \texttt{(sel)}$  &  & \\ \cline{1-3}


     \texttt{\tableop{Reassemble}<R,SEL,a,b> (in: [Strm<R,a>], } & $[D^s_b,\cdots,D^s_0]\ \texttt{(sel)}$ & 
     $[D^s_b,\cdots,D^s_0,$ & \\
     
     \phantom{\texttt{\tableop{Reassemble}<R,SEL,a,b> }} \texttt{sel: Strm<SEL,b>) $\rightarrow$ Strm<R,a+b+1>} & $[D^i_a, D^i_{a-1},\cdots,D^i_0]_i\ \texttt{(in)}$ & \ \ $D^{sel}_a,D_{a-1},\cdots,D_0]$ & \\ \cline{1-3}


     \texttt{\tableop{EagerMerge}<R,SEL,a> (in: [Strm<R,a>])} & $[D^i_a, D^i_{a-1},\cdots,D^i_0]_i$ &  $[\sum_i D^i_a,D_{a-1},\cdots,D_0]\ \texttt{(data)}$ & \\
     
     \phantom{\texttt{\tableop{EagerMerge}<R,SEL,a> (}} \texttt{$\rightarrow$ Strm<R,a>, Strm<SEL,0>} &  &  $[\sum_i D^i_a]\ \texttt{(sel)}$ & \\ \arrayrulecolor{black} \hline

    \end{tabular}
    \caption{STeP routing and merging operators.
    }
    \label{tab:step-route-merge}
\end{table*}

\begin{table*}
    \centering
    \small 
    \setlength{\tabcolsep}{3pt} 
    \begin{tabular}{llll}
    \hline
    \multicolumn{1}{c}{Operator Signature} & In Stream Shape & Out Stream Shape &  \\ \hline
    \arrayrulecolor{lightgray}
    \texttt{\tableop{Flatten}<R,a,min,max> (in: Strm<R,a>) $\rightarrow$ Strm<R,a-(max-min)>} & $[D_a,\cdots,D_{max},\cdots,$ & $[D_a,\cdots,D_{new},\cdots,D_0]$ & \\ 
     & \ $D_{min},\cdots,D_0]$ &  $(D_{new}=\Pi_{i=min}^{max}D_i)$ & \\ \cline{1-3}

     \texttt{\tableop{Reshape}<A,a,b> (in: Strm<A,a>, chunk\_size: int,} & $[D_a,\cdots,D_b,$ & $[D_a\cdots,\left\lfloor \frac{(D_b+S-1)}{S} \right\rfloor,S,$ & \\ 
     
     \phantom{\texttt{\tableop{Reshape}<A,a,b> }} \texttt{pad: Option<A>) } & \ \ $D_{b-1}\cdots,D_0]$ & \ \ $D_{b-1}\cdots,D_0]_\texttt{ (data, padding)}$ & \\ 
     
     \phantom{\texttt{\tableop{Reshape}<A,a,b> }}\texttt{$\rightarrow$ Strm<A,a+1>, Strm<bool,a+1>} &  & ($S$ = \texttt{chunk\_size}) & \\ \cline{1-3}

     \texttt{\tableop{Promote}<R,a> (in: Strm<R,a>) $\rightarrow$ Strm<R,a+1>} & $[D_a,\cdots,D_0]$ & $[D_{a+1} ,D_a,\cdots,D_0]$ & \\ 
     
     &  & $(D_{a+1}=(1\ if\ (D_a>0)\ else\ 0))$ & \\ 
     
     \cline{1-3}

     \texttt{\tableop{Expand}<R',R,a> (in: Strm<R',a>, ref: Str<R,a>,)} & $[D_a,\cdots,1_b,\cdots,1_0]_\texttt{ (data)}$ & $[D_a,\cdots,D_b,\cdots,D_0]$ & \\

     \phantom{\texttt{\tableop{Expand}<A,R,a> (}} \texttt{b: int) $\rightarrow$ Strm<R',a>} & $[D_a,\cdots,D_b,\cdots,D_0]_\texttt{ (ref)}$ &  & \\
     
     \cline{1-3}

     \texttt{\tableop{Zip}<R,R',a> (in1: Strm<R,a>, in2: Str<R$'$,a>)$\rightarrow$Strm<(R,R$'$),a>} & $[D_a,\cdots,D_0]_\texttt{ (in1,in2)}$ & $[D_a,\cdots,D_0]$ & \\
    \arrayrulecolor{black} \hline
    \end{tabular}
    \caption{STeP shape operators.
    }
    \label{tab:step-shape-ops}
\end{table*}

\subsection{Hierarchical Tiling}\label{subsec:hierarchical-tiling}
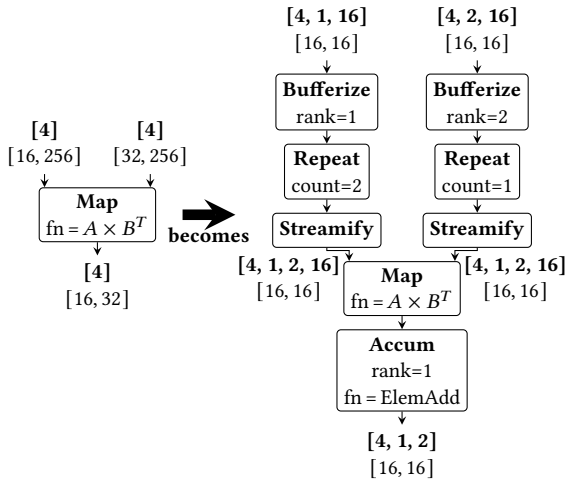
\begin{figure}[b]
    \centering  
    \begin{tikzpicture}
\footnotesize
    \node[rounded corners=1.5pt, draw, align=center] (bigmap) at (-1, 0.5) {
    \textbf{Map}\\
    fn\,=\,$A\times B^T $
    };
    
    \draw[stealth-] ([yshift=-5]bigmap.south) node [yshift=0, below, anchor=north, align=center] {
    \textbf{[4]}\\ 
     $[16, 32]$
    } -- (bigmap);
    
    \draw[-stealth] ([yshift=5, xshift=-20]bigmap.north) -- ([xshift=-20]bigmap.north) node[yshift=5, above, anchor=south, align=center] {
    \textbf{[4]}\\ 
     $[16, 256]$
    };
    \draw[-stealth] ([yshift=5, xshift=20]bigmap.north) -- ([xshift=20]bigmap.north) node [yshift=5, above, anchor=south, align=center] {
    \textbf{[4]}\\ 
     $[32, 256]$
    };

    \draw[-stealth, line width=4pt] ([xshift=10]bigmap.east) -- node[midway, anchor=north] {\textbf{becomes}} ([xshift=30]bigmap.east);

\begin{scope}[shift={(.3, 1cm)}]

    \node [rounded corners=1.5pt, draw, align=center] (buf_1) at (1.75, 1) {\textbf{Bufferize}\\rank=1};
    \node [rounded corners=1.5pt, draw, align=center, anchor=north] (rpt_1) at ([yshift=-5pt]buf_1.south) {\textbf{Repeat}\\count=2};
    \node [rounded corners=1.5pt, draw, align=center, anchor=north] (str_1) at ([yshift=-5pt]rpt_1.south) {\textbf{Streamify}};

    \draw[-stealth] (buf_1) -- (rpt_1);
    \draw[-stealth] (rpt_1) -- (str_1);

    \node [rounded corners=1.5pt, draw, align=center] (buf_2) at (3.75, 1) {\textbf{Bufferize}\\rank=2};
    \node [rounded corners=1.5pt, draw, align=center, anchor=north] (rpt_2) at ([yshift=-5pt]buf_2.south) {\textbf{Repeat}\\count=1};
    \node [rounded corners=1.5pt, draw, align=center, anchor=north] (str_2) at ([yshift=-5pt]rpt_2.south) {\textbf{Streamify}};

    \draw[-stealth] (buf_2) -- (rpt_2);
    \draw[-stealth] (rpt_2) -- (str_2);

    \node[rounded corners=1.5pt, draw, align=center, anchor=north] (map) at ([yshift=-5pt, xshift=1cm]str_1.south) {
    \textbf{Map}\\
    fn\,=\,$A\times B^T $
    };

    \draw[-stealth] (str_1.south) -- ([yshift=-1pt]str_1.south) -| ([xshift=-20pt]map.north);
    \draw[-stealth] (str_2.south) -- ([yshift=-1pt]str_2.south) -| ([xshift= 20pt]map.north);

    \node[rounded corners=1.5pt, draw, align=center, anchor=north] (accum) at ([yshift=-5pt]map.south) {
    \textbf{Accum}\\
    rank=1\\
    fn\,=\,ElemAdd
    };

    \draw[-stealth] (map) -- (accum);
    \draw[-stealth] (accum) -- ([yshift=-5pt]accum.south) node[yshift=0, below, anchor=north, align=center] {
    \textbf{[4, 1, 2]}\\ 
     $[16, 16]$
    };
    \draw[-stealth] ([yshift=5pt]buf_1.north) -- (buf_1.north) node[yshift=5, above, anchor=south, align=center] {
    \textbf{[4, 1, 16]}\\ 
     $[16, 16]$
    };
    \draw[-stealth] ([yshift=5pt]buf_2.north) -- (buf_2.north) node[yshift=5, above, anchor=south, align=center] {
    \textbf{[4, 2, 16]}\\ 
     $[16, 16]$
    };

    \node[rounded corners=1.5pt, align=center, anchor=north west] (str_1_annot) at ([yshift= 0pt, xshift=-16pt]str_1.south west) {\textbf{[4, 1, 2, 16]}\\$[16,16]$};

    \node[rounded corners=1.5pt, align=center, anchor=north east] (str_2_annot) at ([yshift=-0pt, xshift=16pt]str_2.south east) {\textbf{[4, 1, 2, 16]}\\$[16,16]$};
\end{scope}
\end{tikzpicture}
    \caption{Conversion of STeP $A\times B^T$ map node of large tile size to smaller tile size.}
    \label{fig:downtiling}
    \Description{A Map node executing matrix multiplication is decomposed into Bufferize - Repeat - Streamify - Map - Accum to decompose the computation into the physical tile granularity.}
\end{figure}

When mapping to the HDL simulator described in \Cref{subsec:validation}, we apply hierarchical tiling to the tiles in each stream. The larger logical tiles defined at the STeP level are partitioned into smaller physical tiles that match the fabric's compute tile size. \Cref{fig:downtiling} shows an example graph transformation for hierarchical tiling. As shown in the graph, STeP operators and the shape semantics can also be used to express hierarchical tiling.

\subsection{Dataset}\label{subsec:dataset-appendix}
To create batch data from the AzureLLMInference dataset \cite{splitwise}, we batch multiple requests within a 5,000-request time window and compute the standard deviation of KV cache lengths for each batch. We experiment with batches whose prompt length standard deviation matches that of the full 5,000-requests, and batches with the top 10\% highest and lowest variability. 
 
For the expert routing data in the MoE layers, we run Qwen3-30B-A3B~\cite{qwen3} and Mixtral8x7B~\cite{mixtral} using the real-world serving trace HH-RLHF~\cite{hh-rlhf}. To select representative cases, we measure the standard deviation of expert bin counts across iterations and layers, and choose the one whose deviation is closest to the overall average.

\subsection{Dynamic Tiling}\label{subsec:dyn-tiling-appendix}
\begin{figure}[h]
        \centering
        \includegraphics[width=0.95\columnwidth]{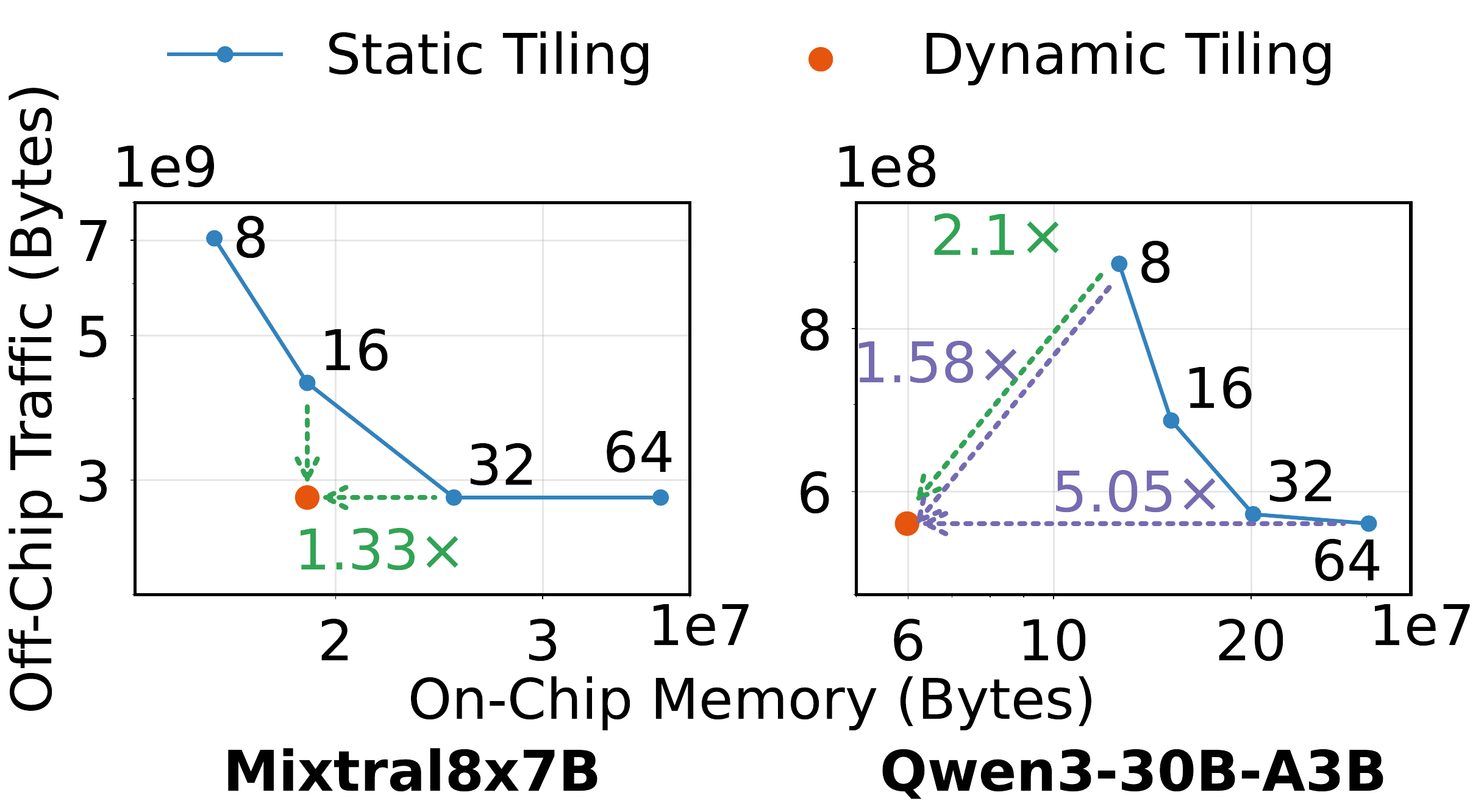}
        \caption{Off-chip traffic and memory requirements of tiling strategies for the batch dimension of each expert (batch = 64). The numbers on the static tiling curve denote tile size. Purple dotted arrows indicate the on-chip memory savings and green dotted arrows indicate speedup.}
        \label{fig:offchip-pareto}
        \Description{A Pareto curve for Mixtral8x7B and Qwen3-30B-A3B with batch size of 64. The y-axis is the off-chip traffic, and the x-axis is the on-chip memory usage.}
    \end{figure}

\begin{figure}[h]
        \centering
        \includegraphics[width=0.95\columnwidth]{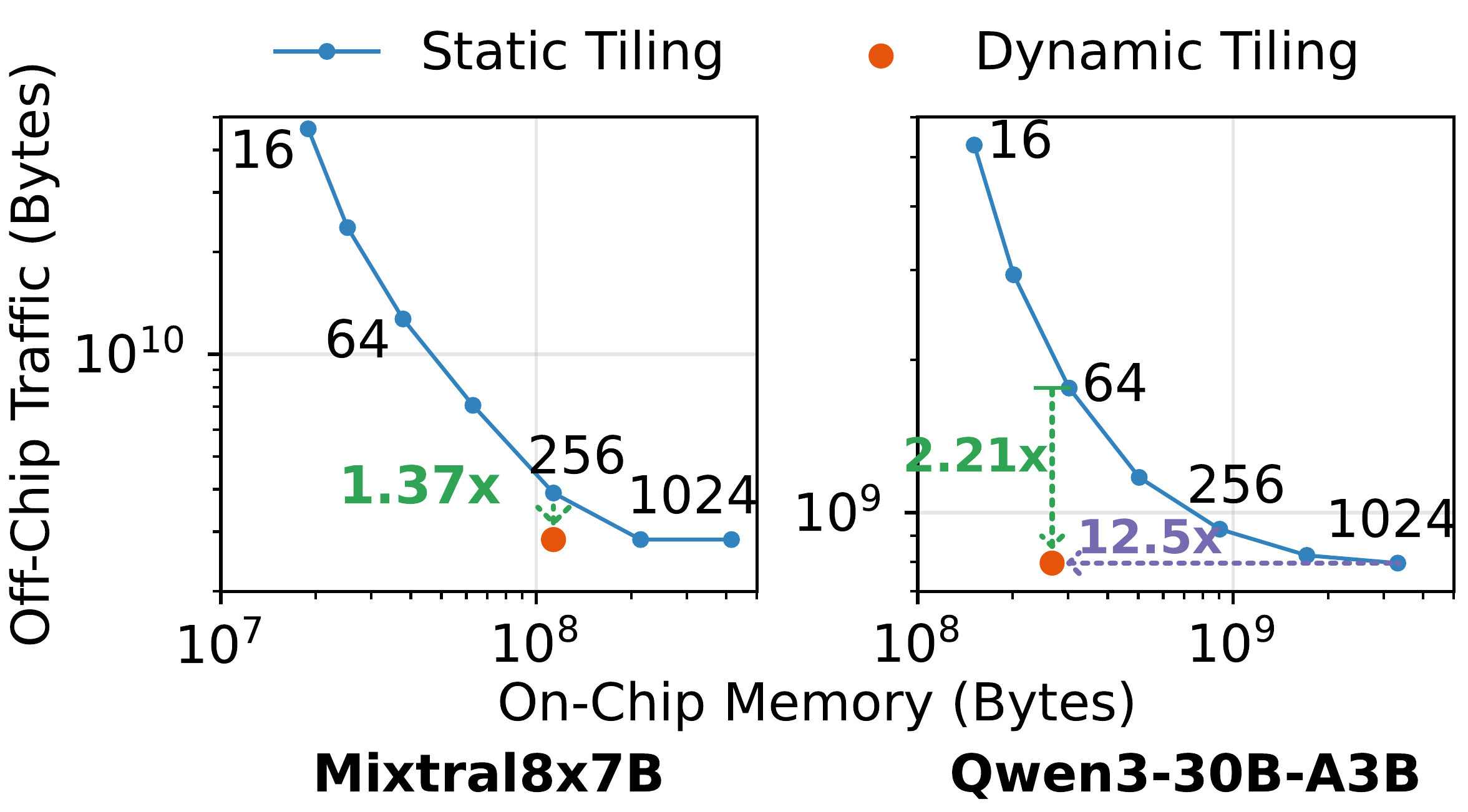}
        \caption{Off-chip traffic and memory requirements of tiling strategies for the batch dimension (batch = 1024). Purple dotted arrows indicate the on-chip memory savings and green dotted arrows indicate speedup.}
        \Description{A Pareto curve for Mixtral8x7B and Qwen3-30B-A3B with batch size of 1024. The y-axis is the off-chip traffic, and the x-axis is the on-chip memory usage.}
        \label{fig:offchip-pareto-1024}
    \end{figure}

In \Cref{subsec:dyn-tiling}, we only show the Pareto curve for performance and on-chip memory requirement, as the performance and off-chip memory traffic show the same trend due to the application being memory-bound in the simulated hardware configuration. The pareto curve for the off-chip memory traffic and on-chip memory usages is in \Cref{fig:offchip-pareto,fig:offchip-pareto-1024}.

To show that dynamic tiling extends the Pareto frontier beyond what is attainable with static tiling, we use the Pareto Improvement Distance metric (PID).
The PID measures the distance from a new design point $p$ to a reference Pareto frontier $F_B$ as the smallest worst-objective multiplicative improvement required for some $q \in F_B$ to match $p$ on all objectives. This can be seen as similar to the single-point specialization of the Average Distance from Reference Set (ADRS) metric~\cite{ferretti2020,meng2016,fist2020}.
We treat both objectives as minimization: cycle count and on-chip memory. Let $F_B$ denote the Pareto-optimal subset of baseline (static) points after removing dominated configurations. For a new point $p$ (e.g., \texttt{tile=dynamic}), we measure its distance to the baseline frontier by comparing it to every $q \in F_B$ and computing the smallest multiplicative factor by which a baseline point would need to improve to match $p$ in \emph{both} objectives simultaneously:
\begin{equation}
\mathrm{PID}(p)
\;=\;
\min_{q \in F_B}\;
\max\!\left(
\frac{\mathrm{cycles}(q)}{\mathrm{cycles}(p)},
\frac{\mathrm{mem}(q)}{\mathrm{mem}(p)}
\right).
\end{equation}
Intuitively, for each baseline frontier point $q$, the inner $\max(\cdot)$ selects the harder objective to match (cycles or memory), and the outer $\min(\cdot)$ picks the baseline point closest to $p$ under this worst-case ratio. This yields a single interpretable number: $\mathrm{PID}(p) > 1$ means $p$ lies strictly beyond the baseline frontier, $\mathrm{PID}(p)=1$ means $p$ lies on the frontier, and $\mathrm{PID}(p)<1$ means $p$ is dominated by the baseline frontier.

\subsection{Dynamic Parallelization}
\begin{figure}[h]
    \centering
    \includegraphics[width=\linewidth]{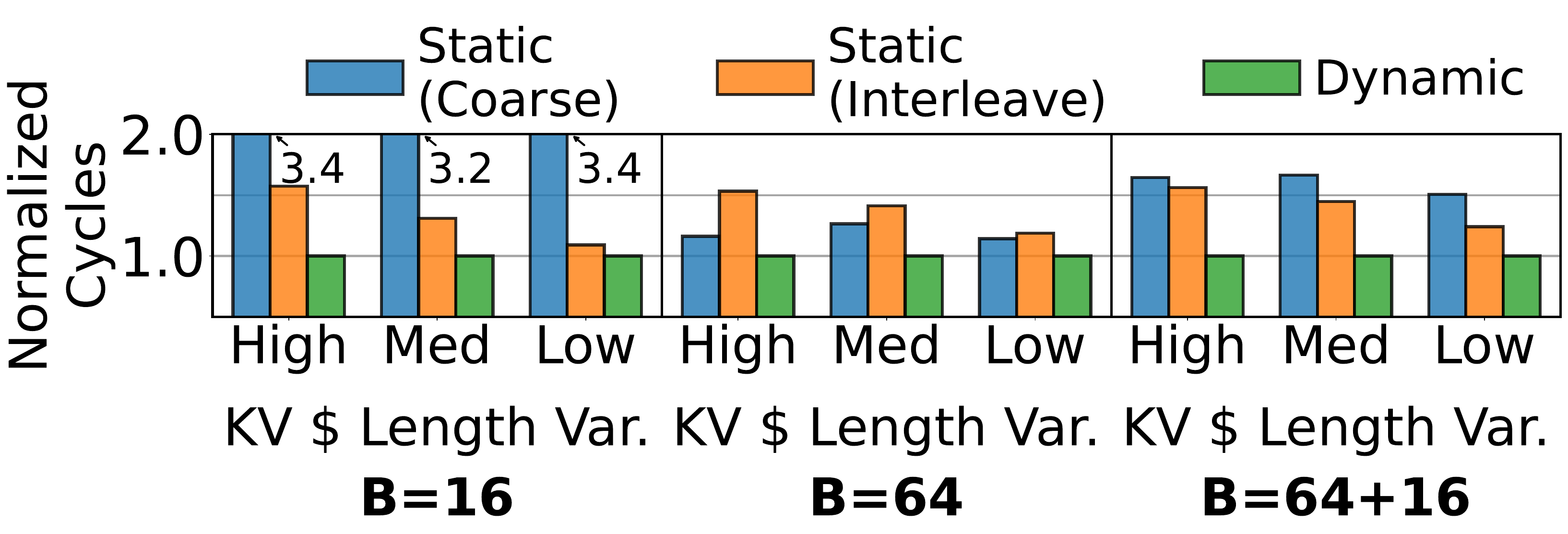}
    \caption{Normalized performance of parallelization strategies relative to dynamic parallelization. For each class, we sample three batches and report the geometric mean performance. KV \$ is used as shorthand for KV cache. 
    }
    \Description{A grouped bar plot comparing the normalized performance (cycles) of static coarse-grained parallelization, static interleaved parallelization, and dynamic parallelization. It sweeps three different batch sizes (16, 64, 80) and batches with low, median, and high KV cache length variation.}
    \label{fig:dyn-par-result}
\end{figure}

As shown in \Cref{fig:dyn-par-result}, static interleave parallelization performs better for smaller batch sizes because the coarse-grained static parallelization can only utilize a portion of the allocated resource when receiving smaller batch sizes.
However, for larger batch sizes, static coarse-grained parallelization performs better as it avoids workload distribution being blocked by a single request with a long KV cache. To avoid this blocking, the interleaving static parallelization requires large buffers in front of each parallel region. However, static coarse-grained parallelization will still suffer from the load imbalance across parallel regions. We also simulate the case where a batch with size 64 and 16 is pipelined as micro batches to see the aggregate effect under different batch sizes. Dynamic parallelization consistently outperforms static parallelization across different batch sizes and KV cache length distributions by dispatching work to parallel regions as soon as they become free. 
Overall, static interleaved parallelization and static coarse-grained parallelization achieve geometric mean speedups of 1.36× and 1.85×, respectively.

\end{document}